\documentclass[aps,prd,preprint,superscriptaddress,floatfix,nobibnotes,nofootinbib]{revtex4-2}

\usepackage{verbatim} 
\usepackage{latexsym}
\usepackage{amsmath}
\usepackage{amssymb}
\usepackage{graphicx}
\usepackage{longtable}

\usepackage[normalem]{ulem}
\usepackage[usenames, dvipsnames]{color}
\usepackage[colorlinks=true]{hyperref}
\usepackage{dsfont}
\usepackage{setspace}
\usepackage{float}

\newcommand{\be}{\begin{equation}}
\newcommand{\ee}{\end{equation}}
\newcommand{\ba}{\begin{eqnarray}} 
\newcommand{\ea}{\end{eqnarray}} 
\newcommand{\nn}{\nonumber}

\numberwithin{equation}{section}

\DeclareMathAlphabet\mathbfcal{OMS}{cmsy}{b}{n}

\begin{document} 

\title{Glasma properties in small proper time expansion}

\author{Margaret E. Carrington}
\affiliation{Department of Physics, Brandon University,
Brandon, Manitoba R7A 6A9, Canada}
\affiliation{Winnipeg Institute for Theoretical Physics, Winnipeg, Manitoba, Canada}

\author{Wade N. Cowie}
\affiliation{Department of Physics, Brandon University,
Brandon, Manitoba R7A 6A9, Canada}

\author{Bryce T. Friesen}
\affiliation{Department of Physics, Brandon University,
Brandon, Manitoba R7A 6A9, Canada}

\author{Stanis\l aw Mr\' owczy\' nski} 
\affiliation{National Centre for Nuclear Research, ul. Pasteura 7,  PL-02-093 Warsaw, Poland}
\affiliation{Institute of Physics, Jan Kochanowski University, ul. Uniwersytecka 7, PL-25-406 Kielce, Poland}

\author{Doug Pickering}
\affiliation{Department of Mathematics, Brandon University,
Brandon, Manitoba R7A 6A9, Canada}

\date{October 20, 2023}

\begin{abstract}

In a series of works by two of us, various characteristics of the glasma from the earliest phase of relativistic heavy-ion collisions have been studied using a proper time expansion. These characteristics include: energy density, longitudinal and transverse pressures, collective flow, angular momentum and parameters of jet quenching. In this paper we extend the proper time interval where our results are reliable by working at higher order in the expansion. We also generalize our previous study of jet quenching by extending our calculations to consider inhomogeneous glasma. Inhomogeneities are an important aspect of physically realistic systems that are difficult to include in calculations and are frequently ignored. 

\end{abstract}

\maketitle
\newpage

\section{Introduction}
\label{sec-intro}

The earliest phase of relativistic heavy-ion collisions is the least understood. The phenomena occurring during this earliest phase are largely `forgotten' due to the subsequent temporal evolution of the system, and consequently experiments provide very limited information about its properties.  What happens during this phase is usually parametrized using a few of its primary characteristics, such as an energy density profile, and used only to provide initial conditions for the later, much better understood, hydrodynamic phase. The earliest phase, however, is of special interest for several reasons. At very early times the matter is strongly anisotropic, far from thermodynamic equilibrium, and the energy density reaches its maximal values. The processes which take place in this phase can significantly affect the subsequent evolution of the system and its final-state characteristics. 

Several different strategies have been used to understand and describe the earliest phase of relativistic heavy-ion collisions. The framework of the Color Glass Condensate effective theory (see, for example, the review \cite{Gelis:2010nm}) is very commonly applied. The theory is based on a separation of scales between hard valence partons and soft gluons. The system that exists at very early times is called a `glasma'. It consists of large occupation number, coherent chromodynamic fields that are essentially classical. The dynamics of the glasma fields is determined by the classical Yang-Mills equations with sources provided by the valence partons. To calculate observables one performs averaging over a Gaussian distribution of colour charges within each nucleus.

Properties of the glasma have been studied for over two decades using more and more advanced numerical simulations, see Refs. \cite{Sun:2019fud,Boguslavski:2021buh,Ipp:2021lwz,Avramescu:2023qvv} as examples of recent works in this direction. There are also analytic approaches, but they are usually very limited in their applicability. There is a method designed to study the earliest phase of relativistic heavy-ion collisions that uses an expansion of the Yang-Mills equations in powers of the proper time $\tau$, which is treated as a small parameter. The method, which is sometimes called a `near field expansion', was proposed in \cite{Fries:2005yc} and further developed in \cite{Fukushima:2007yk,Fujii:2008km,Chen:2015wia,Fries:2017ina,Li:2017iat}. The results provided by the method are limited to small values of $\tau$ but they are analytic and free of artifacts of numerical computation like those caused by taking a continuous limit in the case of lattice calculations. 

The small $\tau$ expansion has been extensively used in a series of works by two of us \cite{Carrington:2020ssh,Carrington:2021qvi,Carrington:2020sww,Carrington:2022bnv,Carrington:2021dvw}. We will summarize below the main results from those papers. The results that will be presented in this paper extend the range of proper times over which our method is reliable, and clarify some important issues about the effects of realistic color charge densities. These results more firmly establish the validity of our method, and motivate the development of other applications. 

In Refs.~\cite{Carrington:2020ssh,Carrington:2021qvi} we studied various glasma characteristics which can be derived from the energy-momentum tensor obtained working to sixth order in the small $\tau$ expansion. The calculation was technically difficult because event-averaged field correlators need to be regulated in both infrared and ultraviolet domains, and because the number and complexity of the terms involved grows rapidly with the order of the proper time expansion. We obtained analytic expressions for the energy density and longitudinal and transverse pressures as functions of $\tau$. Numerical calculations were done using a Woods-Saxon distribution of color charges in the colliding nuclei. Central and peripheral Pb-Pb and Pb-Ca collisions were considered. We discussed the glasma pressure anisotropy and observed the temporal evolution of the longitudinal and transverse pressures, and of two pressures transverse to the beam direction, one parallel to the impact parameter and one perpendicular to it. The beginning of the process of the system's equilibration is clearly seen. 

We also studied in \cite{Carrington:2021qvi} the collective flow of the glasma. We considered radial flow, at fixed azimuthal angle in the transverse plane, and also azimuthal asymmetries of the flow at fixed radius. We found the rather surprising result that Fourier coefficients of flow anisotropy $v_1, v_2, v_3$ are of comparable values to the coefficients experimentally measured in final states of relativistic heavy-ion collisions. We also showed that the glasma collective flow is correlated with the spatial eccentricity of the system which mimics hydrodynamic behavior. The result might explain, at least partially, the success of hydrodynamic models applied to the far from equilibrium quark-gluon plasma that is produced in a heavy ion collision. 

The final finding presented in \cite{Carrington:2021qvi} is a very small glasma angular momentum perpendicular to the reaction plane. This shows that only a small fraction of the very large angular momentum of the incoming nuclei which is carried by the valence quarks, is transferred to the matter produced at midrapidity. Our finding contradicts the picture of a rapidly rotating glasma but it agrees with the experimentally observed absence of global polarization of hyperons and vector mesons produced in heavy-ion collisions at top  RHIC energies and higher \cite{STAR:2017ckg}.

In Refs.~\cite{Carrington:2020sww,Carrington:2022bnv,Carrington:2021dvw} we studied jet quenching in the glasma. Since the preequilibrium phase lasts for about $1~{\rm fm}/c$ or less while the lifetime of the equilibrium phase is an order of magnitude longer, the effect of jet suppression in glasma is typically ignored completely. We have shown that the coefficient $\hat{q}$, which controls the radiative energy loss of a high $p_T$ parton, is about an order of magnitude bigger in glasma than in a quark-gluon plasma in the equilibrium phase. The effect is mostly due to the high energy density of the glasma. Consequently, the accumulated energy loss in the short lifetime preequilibrium phase and in the long lasting equilibrium one are of similar value. The conclusion is that ignoring the glasma in a theoretical description of jet suppression is unjustified. 

In this work we have extended and generalized the calculations done in \cite{Carrington:2020ssh,Carrington:2021qvi,Carrington:2020sww,Carrington:2022bnv,Carrington:2021dvw}. One important point is that the validity of the results obtained using the small proper time expansion depends crucially on the order of the expansion. The properties of the glasma which are derived from the energy-momentum tensor in Refs.~\cite{Carrington:2020ssh,Carrington:2021qvi} were obtained to sixth order. These results hold for $0 < \tau \lesssim 0.05~{\rm fm}/c$. Jet quenching was studied in Refs. \cite{Carrington:2022bnv,Carrington:2021dvw} to fifth order but the radius of convergence is bigger in this case and our results hold for $0 < \tau \lesssim 0.07~{\rm fm}/c$. In this paper we work up to eighth order in the $\tau$ expansion, which extends the interval of $\tau$ where results are reliable. In section \ref{sec-results} we will show that at eighth order in the expansion, results obtained from the energy momentum tensor are valid to approximately $0.06~{\rm fm}/c$, and the momentum broadening coefficient $\hat{q}$ we have calculated is reliable to about $0.08~{\rm fm}/c$. 

The calculation of the chromoelectric and chromomagnetic field correlators at high orders in the proper time expansion is challenging in terms of both computation time and memory. When the fields are written as a sum of terms involving pre-collision potentials, the number of terms grows rapidly with the order of the expansion. For example, at order $\tau^8$ the $z$-component of the magnetic field has $15 \, 964 \,128$ terms. In addition, the number of factors of pre-collision potentials in a single term grows with the order of the expansion, which means that the number of two-point functions produced by applying Wick's theorem also grows quickly. The number of possible contractions of 4 factors of pre-collision potentials is 3. For 12 pre-collison potentials there are $10\,395$ possible contractions. A correlator of two field components at order $\tau^8$ is a sum of approximately $2.6 \times 10^{18}$ terms. Each of these terms is then summed over colour indices. 

The calculation clearly requires the use of computer algebra. {\it Mathematica} is powerful and easy to use, but since glasma potentials are SU$(3)$ valued and therefore non-commutative, {\it Mathematica} (which is heavily based on a built in ordering algorithm) is not well suited to these calculations. There is a package called NCAlgebra  but it is too slow to be useful for our purposes. 
For this reason we have developed a hybrid procedure. The first steps are done using {\it Mathematica}. At each order in the proper time expansion, for pair of squared field components, terms that have an even number of potentials from each nucleus are selected and stored in ordered lists. The remainder of the calculation is done in {\it Julia}, a relatively new language that combines the symbolic features and ease of use of {\it Mathematica's} functional programming with the speed of {\it C}. Separate modules are used to apply Wick's theorem and perform the traces over the colour indices. Up to seventh order, the complete result for each correlator of the form $\langle X^i X^j\rangle$,  where $X^i$ and $X^j$ denote a component of either the chromoelectric or the chromomagnetic field, can be computed and stored, running parallel on 32 cores, in about 6 days. The resulting $6\times 6$ symmetric matrix has 21 independent components and requires about 6 GBytes to store. Beyond seventh order the field correlators are too large to store when they are written in the form of sums of products of pre-collision correlators. It is necessary to write the correlators of the pre-collision potentials in terms of charge density functions, and use numerical values for the coupling constant and the infra-red and ultra-violet regulators introduced in this correlator, before summing all terms and storing the final result. This removes our ability to study the dependence of our results on the gradient expansion, and the values of the confinement and saturation scales, beyond seventh order, without complete re-calculation. All field correlators at eighth order can be calculated in 2 weeks and the final expression is about 10 MBytes. 

In addition to extending our calculations to higher orders in the proper time expansion, we have also improved the calculation presented in \cite{Carrington:2020sww,Carrington:2022bnv,Carrington:2021dvw} in a different way that is particularly important. In \cite{Carrington:2020ssh,Carrington:2021qvi} we included the effects of varying nuclear density in our calculation of the energy momentum tensor using a gradient expansion, similar to the method of Ref.~\cite{Chen:2015wia,Fries:2017ina}. The application of the same method in the momentum broadening and collisional energy loss calculations is considerably more difficult. The basic reason is that the calculation of the energy momentum tensor requires one-point correlators while two-point correlators are needed to calculate transport coefficients. For this reason we calculated $\hat q$ and $dE/dx$ in \cite{Carrington:2020sww,Carrington:2022bnv,Carrington:2021dvw} only in the case of a homogeneous glasma where the incoming nuclei are assumed to be infinitely extended and homogeneous in the plane transverse to the beam direction. A realistic modeling of jet quenching in relativistic heavy-ion collisions obviously requires treating nuclei as finite objects of varying density. This issue is particularly important because the Fokker-Planck formalism that we use relies on some assumptions about the approximate translation invariance of the glasma. In this work we have verified that these assumptions are justified. We have modified and extended our calculations of $\hat q$ and $dE/dx$ so that the glasma under consideration is produced in collisions of finite nuclei with a Woods-Saxon density distribution. The field correlators are computed using the first order gradient expansion and to seventh order in the proper time expansion. 

Throughout the paper we use the natural system of units with $c = \hbar = k_B =1$.

\section{Summary of computational method}
\label{sec-method}

We consider a collision of two heavy ions moving with the speed of light towards each other along the $z$-axis and colliding at $t=z=0$. The vector potential of the gluon field is described with the ansatz \cite{Kovner:1995ts} 
\ba
\nn
A^+(x) &=& \Theta(x^+)\Theta(x^-) x^+ \alpha(\tau,\vec x_\perp) ,
\\\label{ansatz}
A^-(x) &=& -\Theta(x^+)\Theta(x^-) x^- \alpha(\tau,\vec x_\perp) ,
\\ \nn
A^i(x) &=& \Theta(x^+)\Theta(x^-) \alpha_\perp^i(\tau,\vec x_\perp)
+\Theta(-x^+)\Theta(x^-) \beta_1^i(x^-,\vec x_\perp)
+\Theta(x^+)\Theta(-x^-) \beta_2^i(x^+,\vec x_\perp) ,
\ea
where the functions $\beta_1^i(x^-,\vec x_\perp)$ and $\beta_2^i(x^+,\vec x_\perp)$ represent the pre-collision potentials, and the functions $\alpha(\tau,\vec x_\perp)$ and $\alpha_\perp^i(\tau,\vec x_\perp)$ give the post-collision potentials. 

In the forward light-cone the vector potential satisfies the sourceless Yang-Mills equations but the sources enter through the boundary conditions that connect the pre-collision and post-collision potentials. The boundary conditions are
\ba
\label{cond1}
\alpha^{i}_\perp(0,\vec{x}_\perp) &=& \alpha^{i(0)}_\perp(\vec{x}_\perp) 
= \lim_{\text{w}\to 0}\left(\beta^i_1 (x^-,\vec{x}_\perp) + \beta^i_2
(x^+,\vec{x}_\perp)\right) ,
\\
\label{cond2}
\alpha(0,\vec{x}_\perp) &=& \alpha^{(0)}(\vec{x}_\perp) = -\frac{ig}{2}\lim_{\text{w}\to 0}\;[\beta^i_1 (x^-,\vec{x}_\perp),\beta^i_2
(x^+,\vec{x}_\perp)] ,
\ea
where the notation $\lim_{\text{w}\to 0}$ indicates that the width of the sources across the light-cone is taken to zero, as the colliding nuclei are infinitely contracted.

We find solutions valid for early post-collision times by expanding the Yang-Mills equations in the proper time $\tau$. Using these solutions we can write the post-collision field-strength tensor, and energy-momentum tensor, in terms of the initial potentials $\alpha(0, \vec x_\perp)$ and $\vec\alpha_\perp(0, \vec x_\perp)$ and their derivatives, which in turn are expressed through the pre-collision potentials $\vec \beta_1(x^-,\vec x_\perp)$ and $\vec \beta_2(x^+,\vec x_\perp)$ and their derivatives. 

The next step is to use the Yang-Mills equations to write the pre-collision potentials in terms of the color charge distributions of the incoming ions. One then averages over a Gaussian distribution of color charges within each nucleus. The average of a product of color charges can be written as a sum of terms that combine the averages of all possible pairs, which is called Wick's theorem. We use the Glasma Graph approximation \cite{Lappi:2017skr} which means that we apply Wick's theorem not to color charges but to gauge potentials. The correlator of two pre-collision potentials from different ions is assumed to be zero as the potentials are not correlated to each other. The building blocks of all physical quantities we study are the correlators for two potentials from the same ion
\be
\label{core5-20}
\delta^{ab} B_n^{ij}(\vec{x}_\perp,\vec y_\perp) \equiv 
\lim_{{\rm w} \to 0}  \langle \beta_{n\,a}^i(x^-,\vec x_\perp) \beta_{n\,b}^j(y^-,\vec y_\perp)\rangle  ,
~~~~~~n=1,~\,2
\ee
and their derivatives. 
In our calculation we use an expression for the correlators (\ref{core5-20}) that has the standard MV form (see \cite{Carrington:2020ssh} for a detailed derivation), but also includes surface density functions that depend on position in the transverse plane, using the method of Ref. \cite{Chen:2015wia}. 
Other analytic forms for these correlation functions have been studied, for example see \cite{Guerrero-Rodriguez:2021ask, Demirci:2023ejg} for expressions applicable to dilute systems.
The surface density functions for the two ions $\mu_1(\vec x_\perp)$ or $\mu_2(\vec x_\perp)$ are a phenomenological input to our calculations, and we have used a Woods-Saxon distribution projected on the plane transverse to the collision axis of the form 
\ba
\mu(\vec x_\perp) \label{def-mu2-2} 
 =  \left(\frac{A}{207}\right)^{1/3}\frac{\bar\mu}{2a\ln(1+e^{R_A/a})} 
\int^\infty_{-\infty} \frac{dz }{1 + \exp\big[(\sqrt{(\vec x_\perp)^2 + z^2} - R_A)/a\big]}\,.
\ea
The parameters $R_A$ and $a$ give the radius and skin thickness of a nucleus of mass number $A$.
We use $a=0.5$ fm and $R_A=A^{1/3}r_0$ fm with $r_0=1.25$ and $A=207$. 
The integral in (\ref{def-mu2-2}) is normalized so $\bar\mu$ is  the value of the charge density at the center of the nucleus. This parameter is related to the saturation scale $Q_s$ and we make the standard choice $\bar\mu = Q_s^2/g^4$. 

All of our results are obtained for the SU(3) gauge group, $g=1$,  saturation scale $Q_s = 2$ GeV and infrared cutoff $m=0.2$ GeV. 
\color{black}

\section{Results}
\label{sec-results}

In the following four subsections we present results for Pb-Pb collisions at different impact parameters  $\vec b$. A space and time position in a glasma is determined by the proper time $\tau$, the space-time rapidity $\eta$ and the transverse vector $\vec R$. The centers of the two infinitely contracted nuclei at the moment of collision are at $\vec R = \vec b/2$ and $\vec R = -\vec b/2$. The vector $\vec R$ is often written in polar coordinates as $\vec R = (R,\phi)$. When the collision is central and $\vec b = 0$, the glasma system has cylindrical symmetry. Since we are mostly interested in the glasma at midrapidity, the space-time rapidity $\eta$ is set to zero for all calculations presented in this paper.

\subsection{Anisotropy}
\label{sec-aniso}

The transverse and longitudinal pressures $p_T, p_L$ and the energy density ${\cal E}$ are defined through the energy-momentum tensor in Minkowski coordinates $T^{\mu \nu}$ as
\be
{\cal E} \equiv T^{00}, ~~~~~~~~~~~ p_T \equiv T^{xx} , ~~~~~~~~~~~ p_L \equiv T^{zz} .
\ee
To observe the temporal evolution of the glasma anisotropy one can use a measure defined as \cite{Jankowski:2020itt}
\ba
\label{AA-def}
A_{TL} \equiv \frac{3(p_T-p_L)}{2p_T+p_L}.
\ea
Initially the energy-momentum tensor is diagonal and $p_T  = {\cal E}_0 = - p_L$ where ${\cal E}_0$ is the initial energy density. At $\tau=0$ the initial value of the anisotropy measure is therefore $6$. As $\tau$ increases we expect it to evolve towards zero as the glasma isotropizes. Since the energy momentum tensor is a local quantity and the colliding nuclei are finite and of varying density in the transverse plane, the measure $A_{TL}$ and its temporal evolution can non-trivially depend on position.   

\begin{figure}
\begin{center}
\includegraphics[width=10cm]{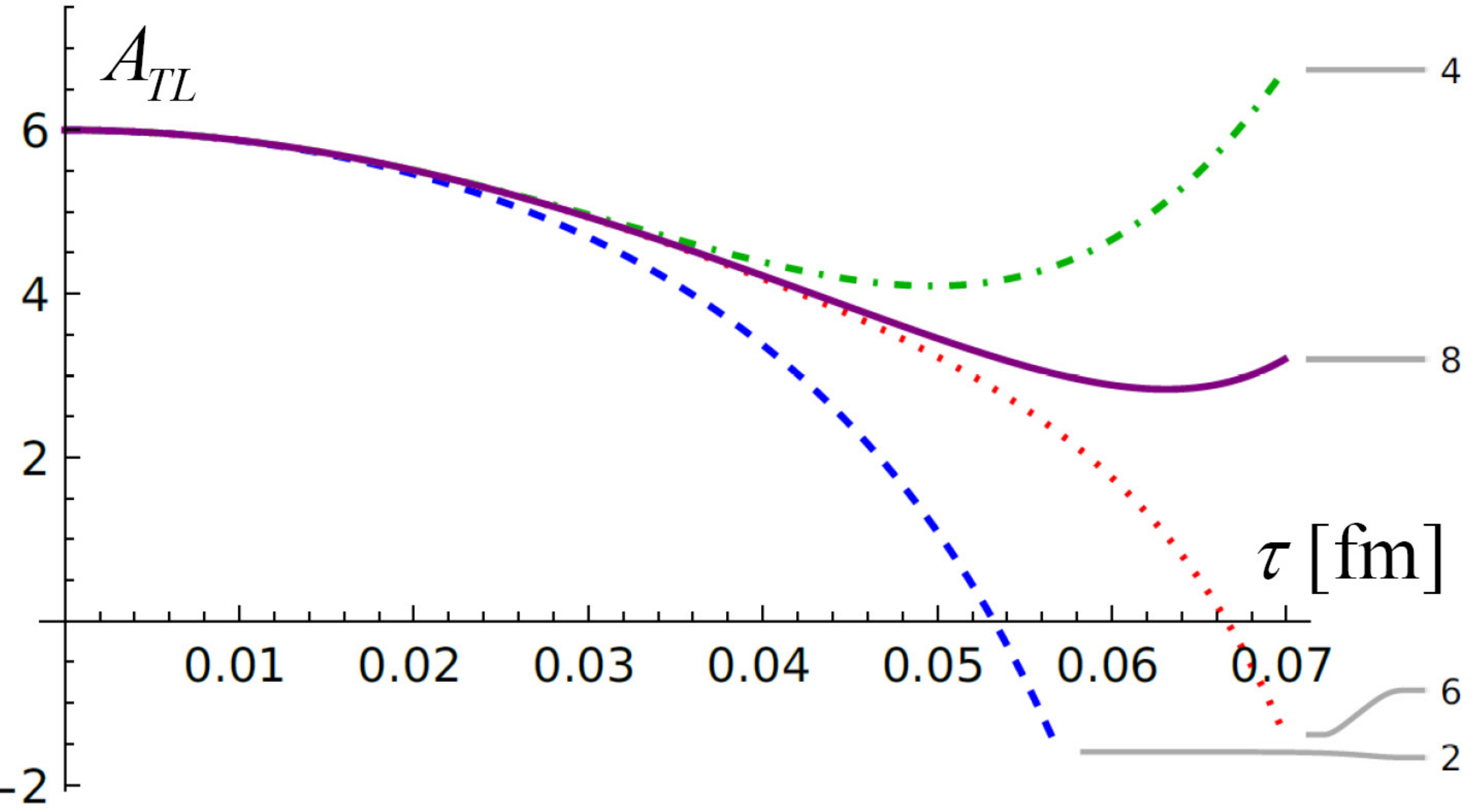}
\end{center}
\vspace{-8mm}
\caption{The anisotropy measure $A_{TL}$ versus $\tau$ at $R = 5$ fm in central collisions ($b=0$).}
\label{figure4-repro}
\end{figure}

In Refs.~\cite{Carrington:2020ssh,Carrington:2021qvi} we worked to sixth order in the proper time expansion and showed that the anisotropy measure $A_{TL}$ decreases as $\tau$ grows up until $\tau \approx 0.05$ fm when the proper time expansion breaks down. Figure \ref{figure4-repro} presents our calculations of central collisions with $b=0$ and $R = 5$ fm to eighth order. One sees that the radius of convergence is extended to about $0.06$ fm. We note that the $n$-th order result is a sum of all contributions up to $n$-th order. 

\begin{figure}
\begin{center}
\includegraphics[width=5.0cm]{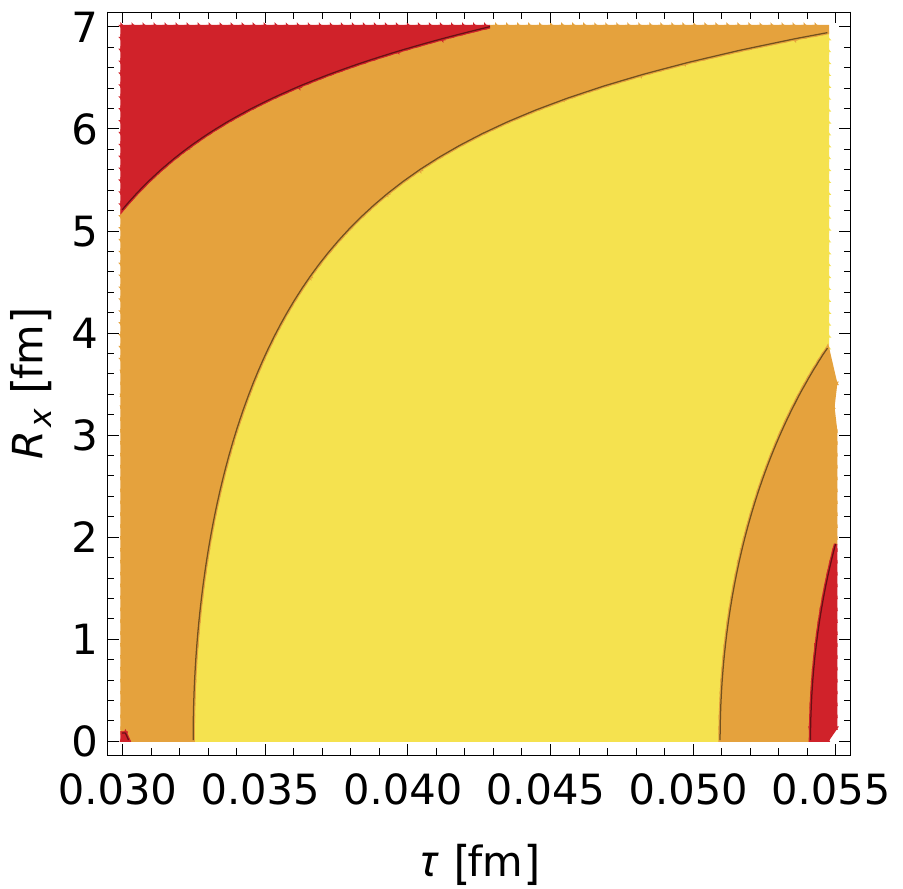}
\includegraphics[width=5.0cm]{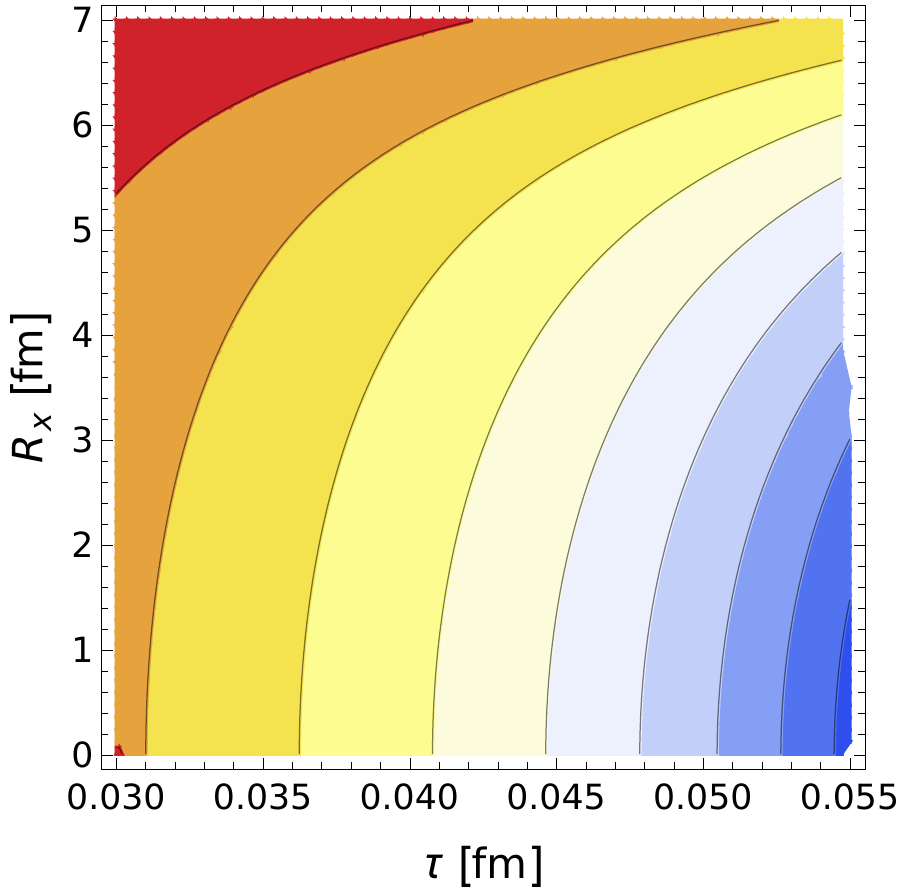}
\includegraphics[width=6.cm]{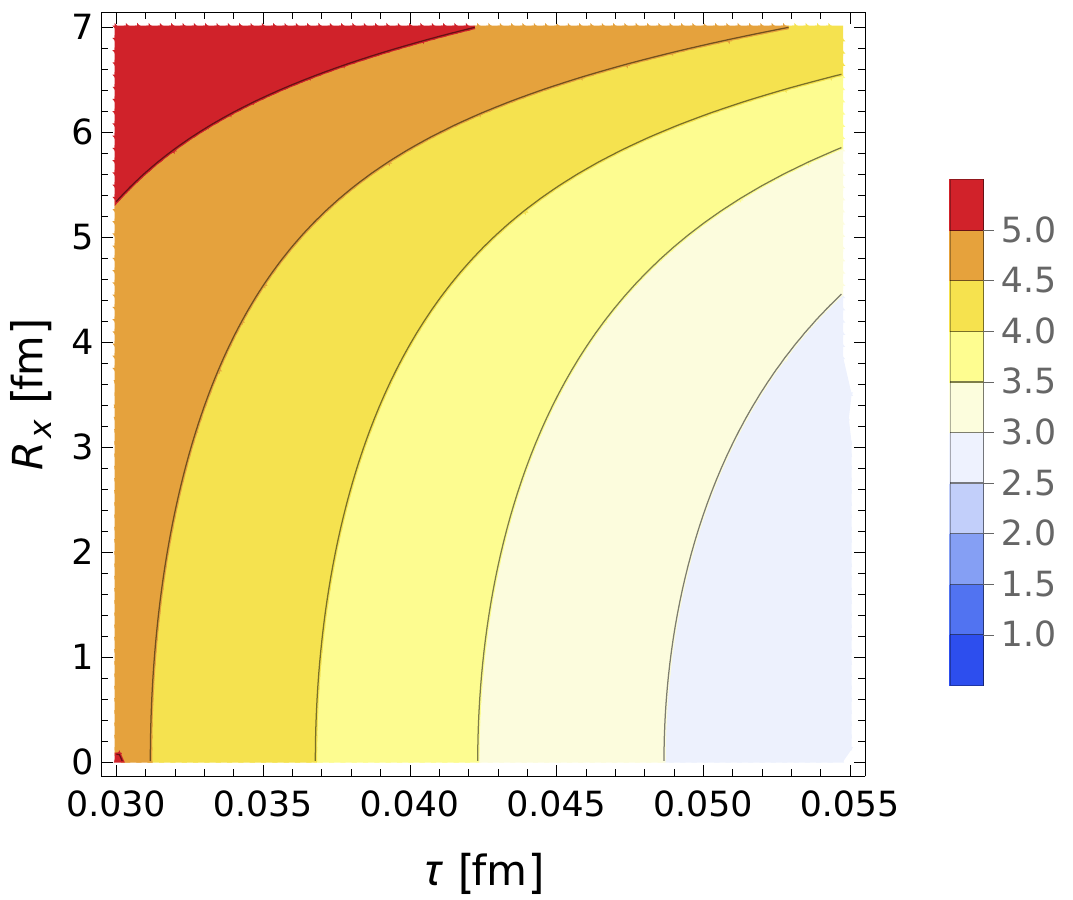}
\end{center}
\vspace{-8mm}
\caption{The anisotropy measure $A_{TL}$ at three different orders in the $\tau$ expansion in central collisions ($b=0$). The left panel shows the result at fourth order in the expansion, the middle is sixth order, and the right panel shows the eighth order result.  See text for further discussion.}
\label{A-R-tau}
\end{figure}

It is interesting to see how the anisotropy measure $A_{TL}$ changes as a function of $R$ and $\tau$ at different orders in the proper time expansion. Figure \ref{A-R-tau} shows contour plots of $A_{TL}$ in central collisions ($b=0$) at different orders in the $\tau$ expansion. The vertical axis corresponds to $R$ in fm and the horizontal axis to $\tau$ in fm. The figure shows that $A_{TL}$ is lowest at the center of the incoming nucleus, and that it decreases with $\tau$ up to the point that the expansion breaks down. There is almost no isotropisation in the outer part of the system. The sixth order result gives a lower minimum $A_{TL}$ but, as can be seen in Fig.~\ref{figure4-repro}, this happens because the sixth order calculation diverges towards negative values when the expansion breaks down. The eighth order result decreases more uniformly across the same range of $R$ and $\tau$, which shows that the system is closer to an isotropic state at eighth order. 

\subsection{Radial flow}
\label{sec-flow}

\begin{figure}
\begin{center}
\includegraphics[width=12cm]{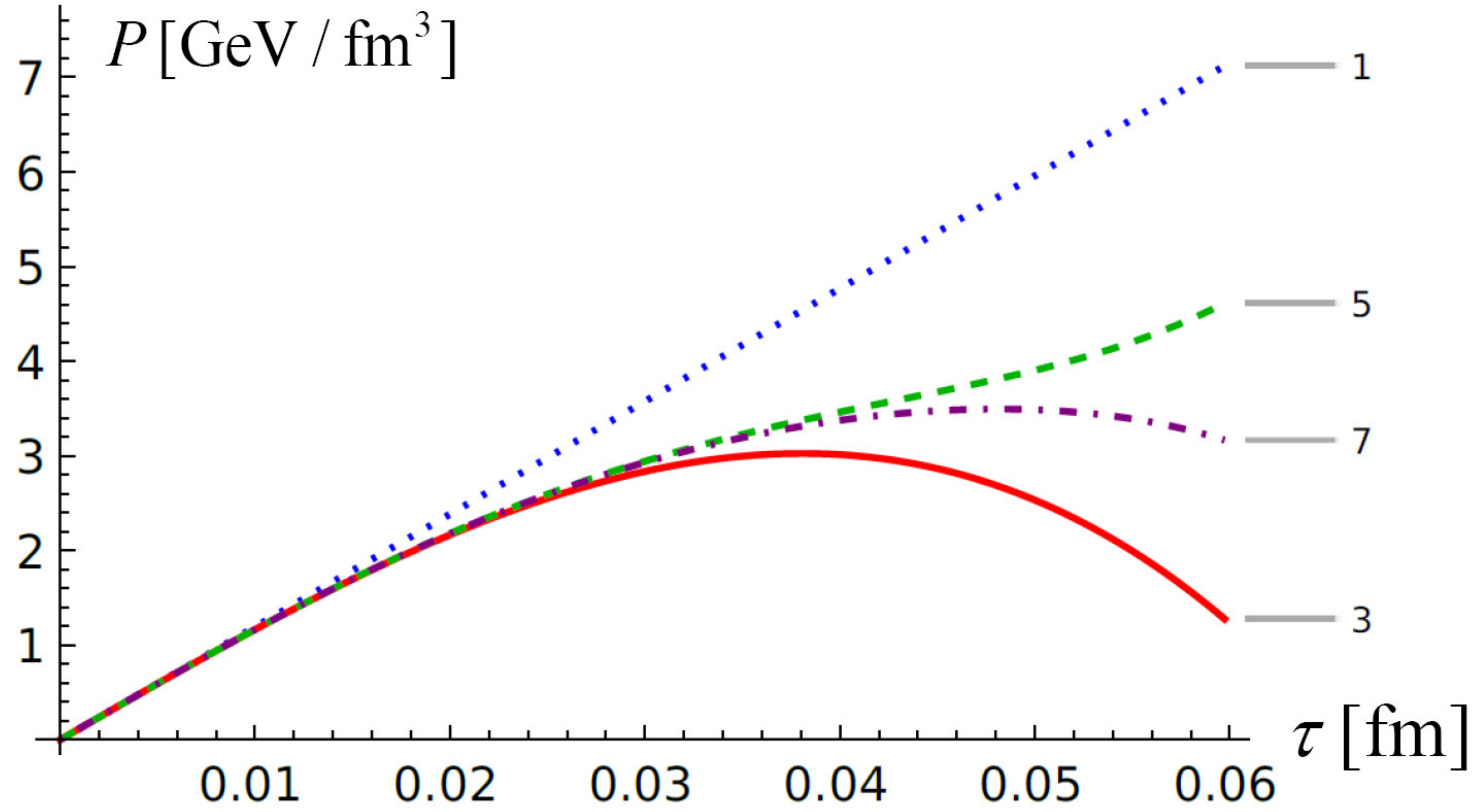}
\end{center}
\vspace{-8mm}
\caption{Radial flow to seventh order in the proper time expansion at $R=3$ fm and $\phi=\pi/2$ in collisions with $b=6$ fm.}
\label{fig-P-time}
\end{figure}

\begin{figure}
\begin{center}
\includegraphics[width=12cm]{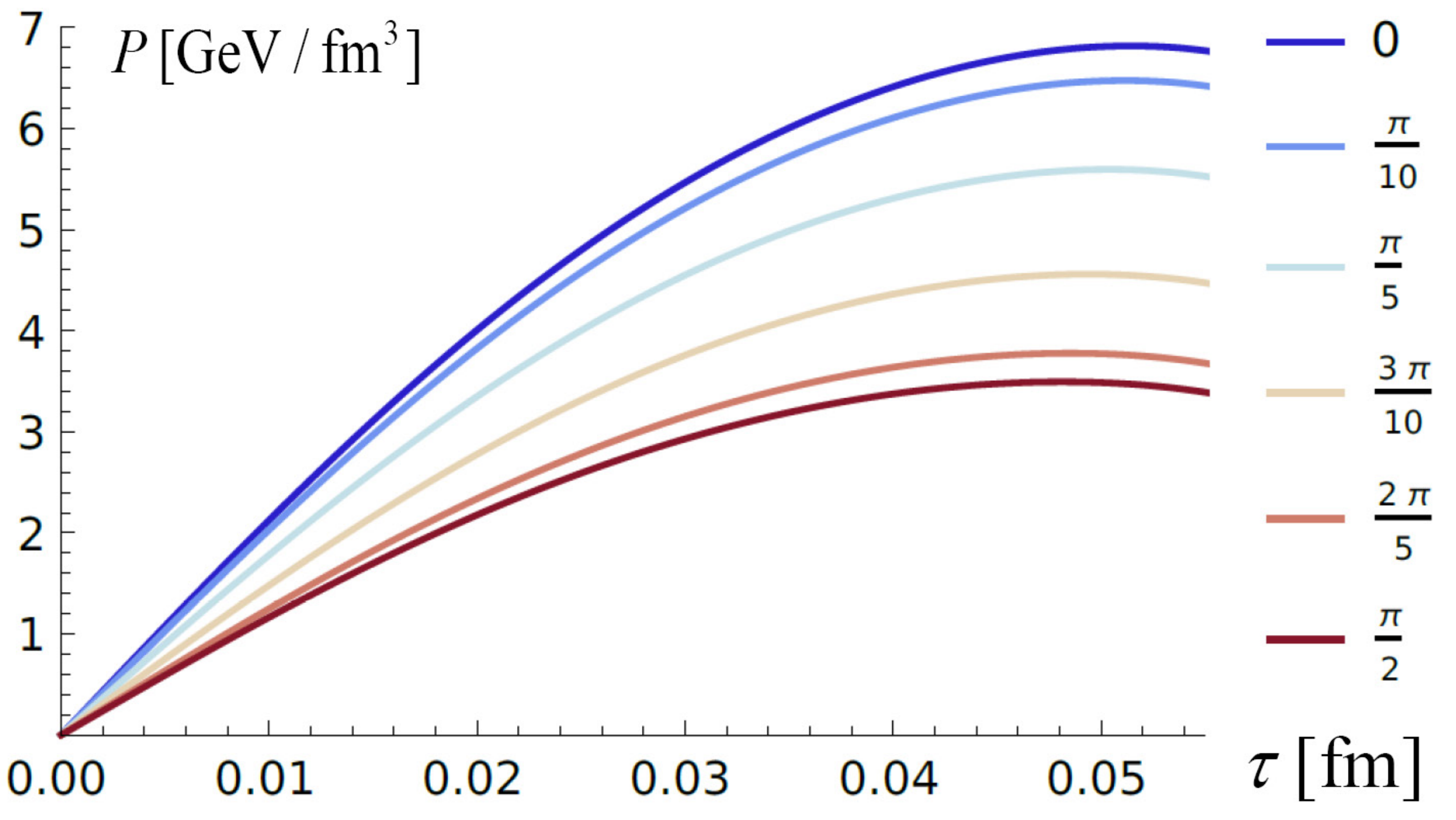}
\end{center}
\vspace{-8mm}
\caption{Radial flow to seventh order in the proper time expansion at $R=3$ fm for a range of azimuthal angles $\phi$ in collisions with $b=6$ fm.}
\label{fig-P-time-phi}
\end{figure}

To characterize the radial flow of the expanding glasma we compute the radial projection of the transverse Poynting vector $P \equiv \hat R^i T^{i0}$ where $\hat R^i \equiv R^i/|\vec{R}|$. In Fig.~\ref{fig-P-time} we show this quantity for fairly peripheral collisions with $b=6$ fm at $R=3$ fm and $\phi=\pi/2$, at different orders in the $\tau$ expansion. Our previous work included only the fifth order contribution. One observes that when seventh order contributions are taken into account, our result for radial flow can be trusted to $\tau \lesssim 0.06$ fm. Figure \ref{fig-P-time-phi} shows the same quantity $P$ at $R=3$ fm for a range of azimuthal angles $\phi$ in collisions with $b=6$ fm. The flow is seen to be significantly stronger in the reaction plane ($\phi=0$) than in the direction perpendicular to it ($\phi=\pi/2$).

\begin{figure}
\begin{center}
\includegraphics[width=7.5cm]{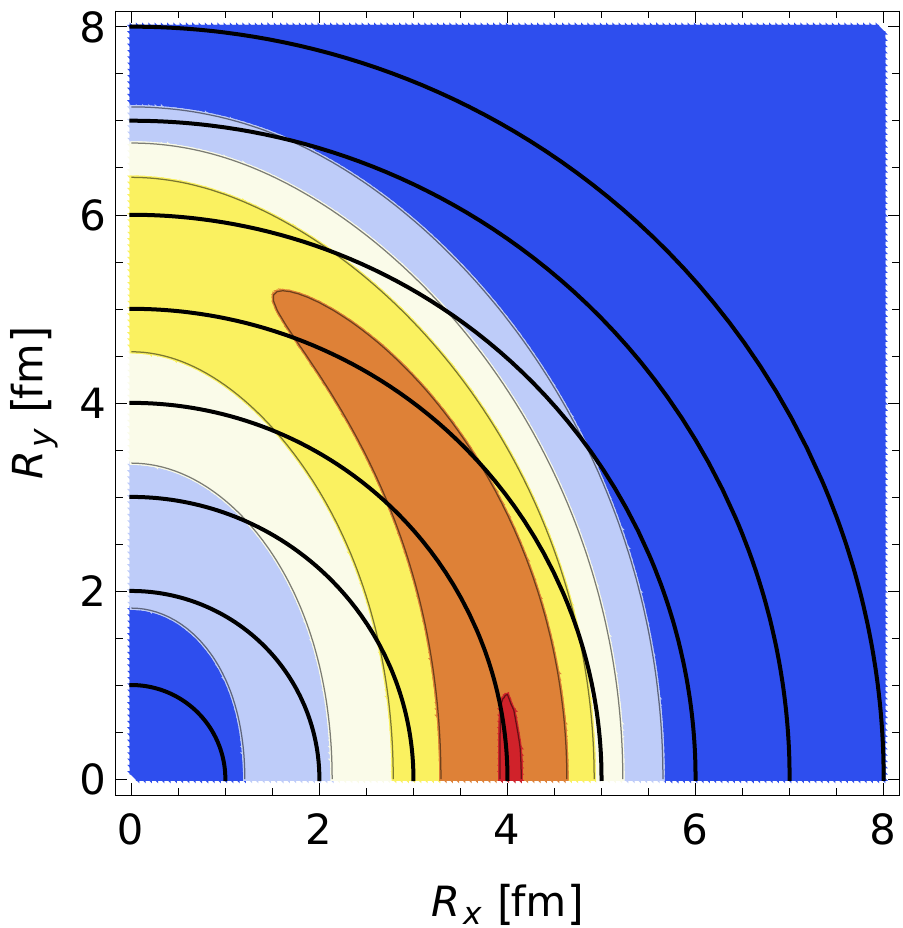}
\includegraphics[width=8.7cm]{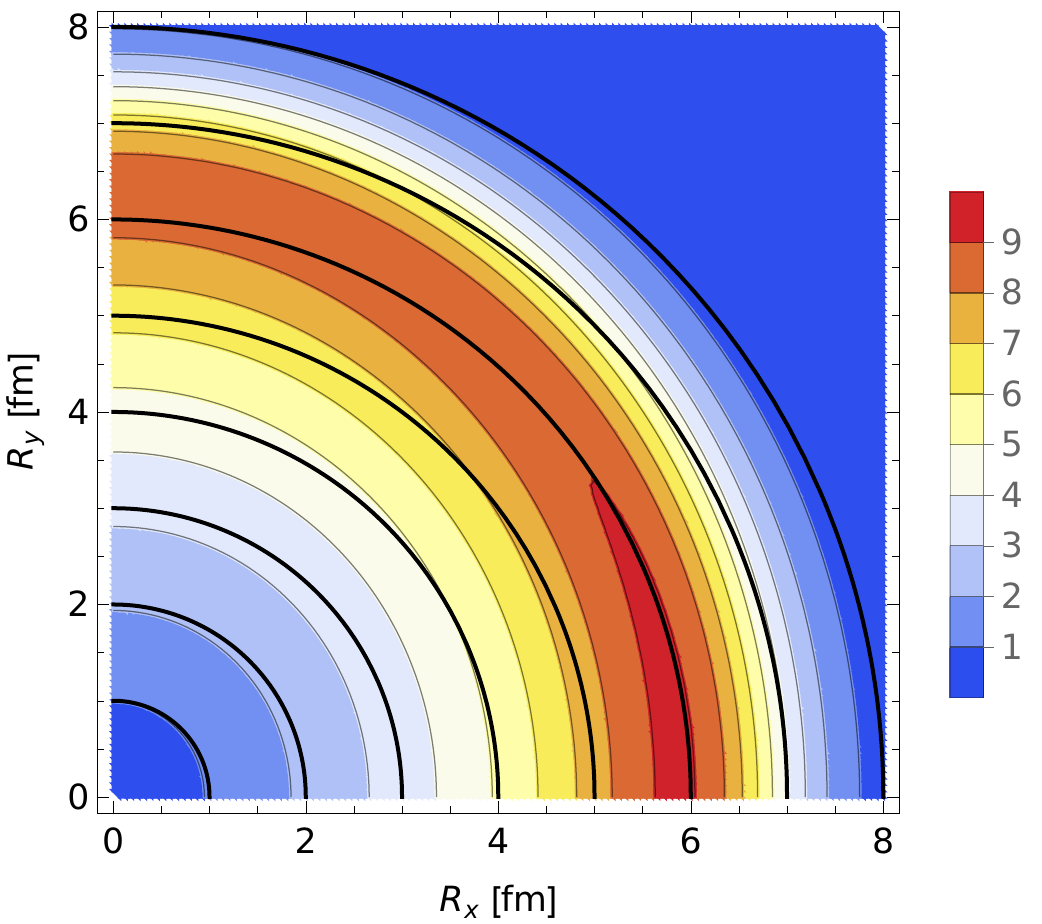}
\end{center}
\vspace{-8mm}
\caption{Radial flow in the transverse plane at $\tau = 0.05$ fm at seventh order in the proper time expansion  in collisions with $b=6$ fm  (left panel) and with $b=2$ fm (right panel). The black curves mark lines of constant radius. See text for further explanation.}
\label{fig-P-trans}
\end{figure}

When the impact parameter is non-zero, we expect that the radial flow in the plane transverse to the beam direction will not be azimuthally symmetric. In our coordinate system the $x$-$y$ plane is transverse to the beam axis, and we always choose the  impact parameter along the $x$-axis. The left panel of Fig.~\ref{fig-P-trans}  shows the radial flow of the glasma for a fairly peripheral collision with $b = 6$ fm, and the right panel is a more central collision with $b = 2$ fm. The flow is greater in the $x$ than in the $y$ direction, as expected, up to $R \approx 5$ fm in the peripheral collision and up to $R  \approx 7$ fm in the more central collision. At bigger distances there is a slight increase in the radial flow at larger azimuthal angles, but since the gradient expansion is not reliable at distances comparable to the nuclear radii, the accuracy of the calculation is much lower in this region. The effect is difficult to see from the figures and the black arcs that represent quarter circles are intended to make it more easily visible.
\subsection{Fourier coefficients of azimuthally asymmetric flow}
\label{sec-Fourier}

The azimuthal asymmetry of the collective flow is usually quantified in terms of Fourier coefficients $v_1,v_2,v_3 \dots$. In Appendix C of our paper \cite{Carrington:2021qvi} we explain in detail how these coefficients are defined and how they are expressed in terms of the components $T^{0x}$ and $T^{0y}$ of the energy-momentum tensor. Below we discuss only the elliptic flow coefficient $v_2$ and the eccentricity of the energy density $\varepsilon$ which are defined as
\be
v_2 = \frac{\int d^2 R \, \frac{T_{0x}^2 - T_{0y}^2}{\sqrt{T_{0x}^2+T_{0y}^2}}}
{\int d^2 R \, \sqrt{T_{0x}^2+T_{0y}^2}}
\text{~~~~~and~~~~~} 
\varepsilon = - \frac{\int d^2 R\,  \frac{R_x^2-R_y^2}{\sqrt{R_x^2+R_y^2}} \,T^{00}}
{\int d^2 R\, \sqrt{R_x^2+R_y^2} \, T^{00}} \,. 
\ee

\begin{figure}
\begin{center}
\includegraphics[width=11cm]{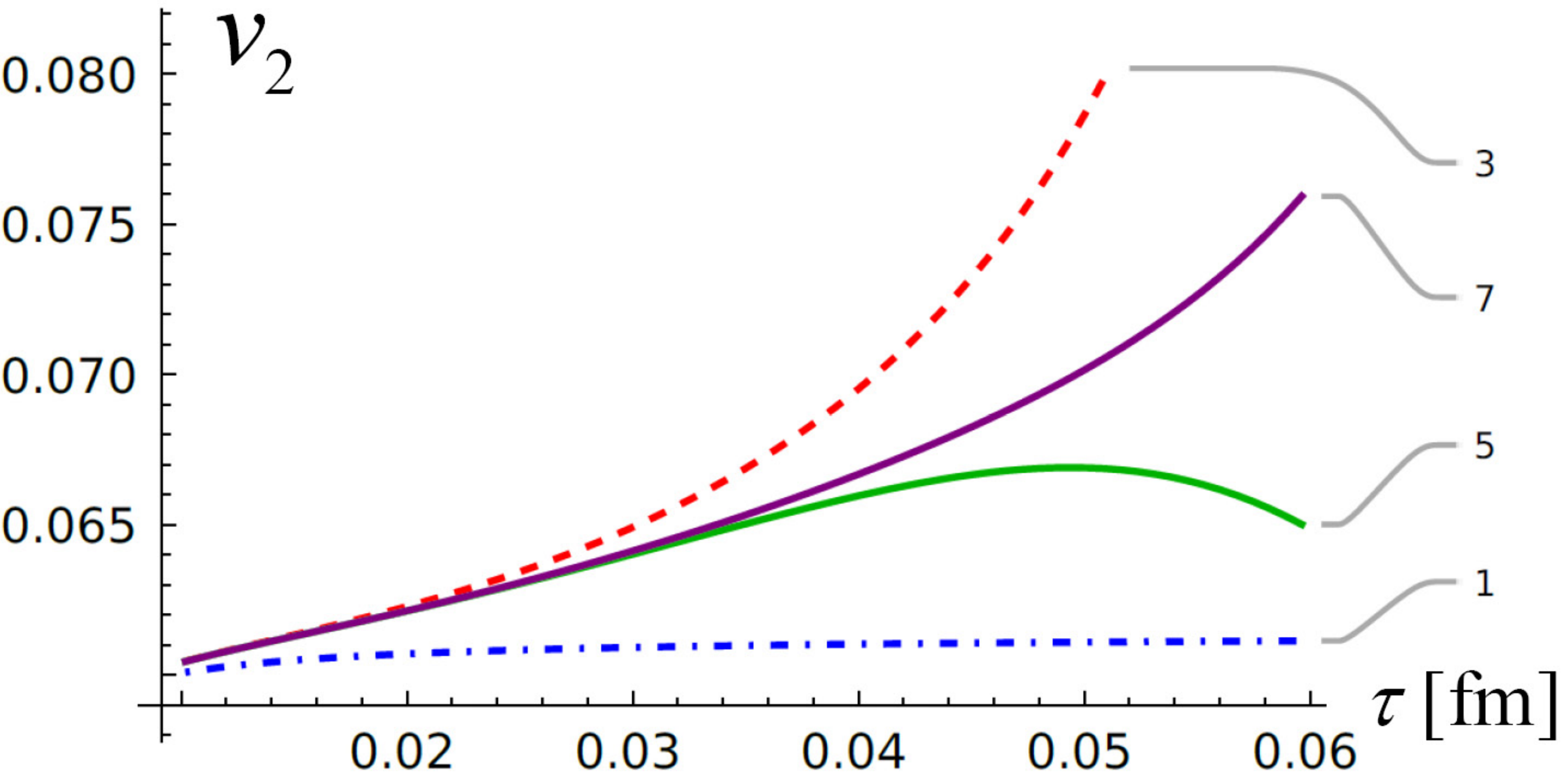}
\end{center}
\vspace{-7mm}
\caption{Elliptic flow coefficient $v_2$ versus proper time at different orders in the proper time expansion in collisions at $b=2$ fm. }
\label{fig-v2-vs-tau}
\end{figure}

In Fig.~\ref{fig-v2-vs-tau} we show the coefficient $v_2$ as a function of $\tau$ at orders one, three, five and seven of the expansion, for collisions with impact parameter $b=2$ fm. The coefficient $v_2$ is constant in time at first order in the expansion, since both the numerator and denominator are linear in $\tau$. The seventh order result clearly shows that $v_2$ does not saturate at $\tau \gtrsim 0.05$ fm, as the fifth order result might suggest, but it continues to grow with time.  We note that the calculation of $v_2$ at very small times is numerically difficult because the numerator and denominator both approach zero as $\tau\to 0$. 

\begin{figure}
\begin{center}
\includegraphics[width=12cm]{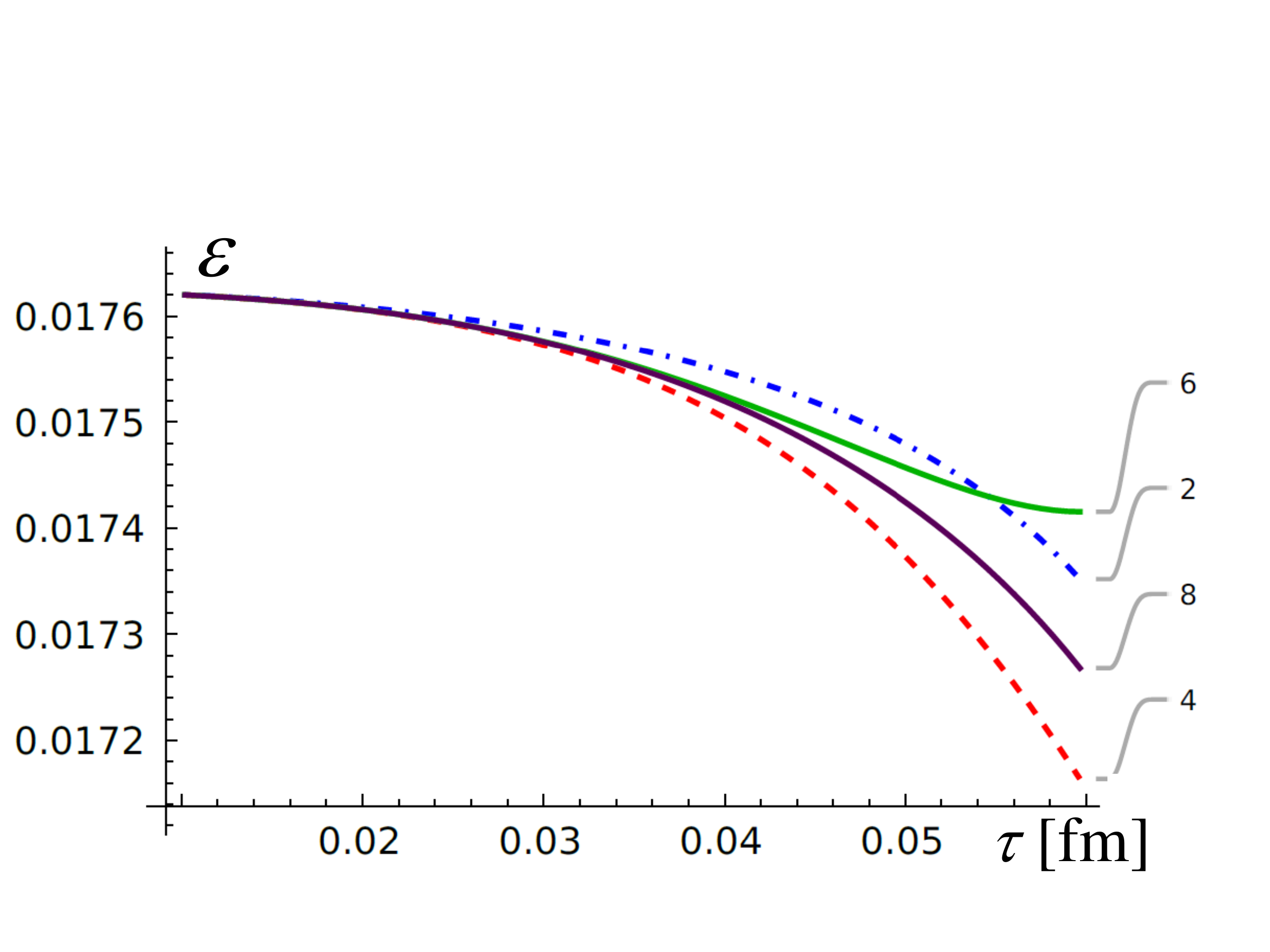}
\end{center}
\vspace{-15mm}
\caption{Eccentricity $\varepsilon$ versus proper time at different orders in the proper time expansion in collisions at $b=2$ fm.}
 \label{fig-e2-vs-tau}
\end{figure}

It is usually assumed that the experimentally observed azimuthal anisotropy in momentum space of a hadronic final state is caused by the azimuthal anisotropy in coordinate space of the energy density and pressure of the initial state. Physically the idea is that the final state momentum anisotropy is generated by pressure gradients, and it is expected that this takes place mostly during the hydrodynamic evolution of the system \cite{Heinz:2013th}. To investigate if this behaviour is seen in our calculation, we have computed the eccentricity $\varepsilon$ as a function of $\tau$ at orders two, four, six and eight of the expansion, for collisions with impact parameter $b=2$ fm. The results are presented in Fig.~\ref{fig-e2-vs-tau} and together with Fig.~\ref{fig-v2-vs-tau} they show that the collective elliptic flow increases in time while the spatial eccentricity decreases, which resembles hydrodynamical behaviour even though the glasma is far from a local equilibrium state. One sees also that the eccentricity changes much more slowly than the elliptic flow coefficient.

\begin{figure}
\begin{center}
\includegraphics[width=11.5cm]{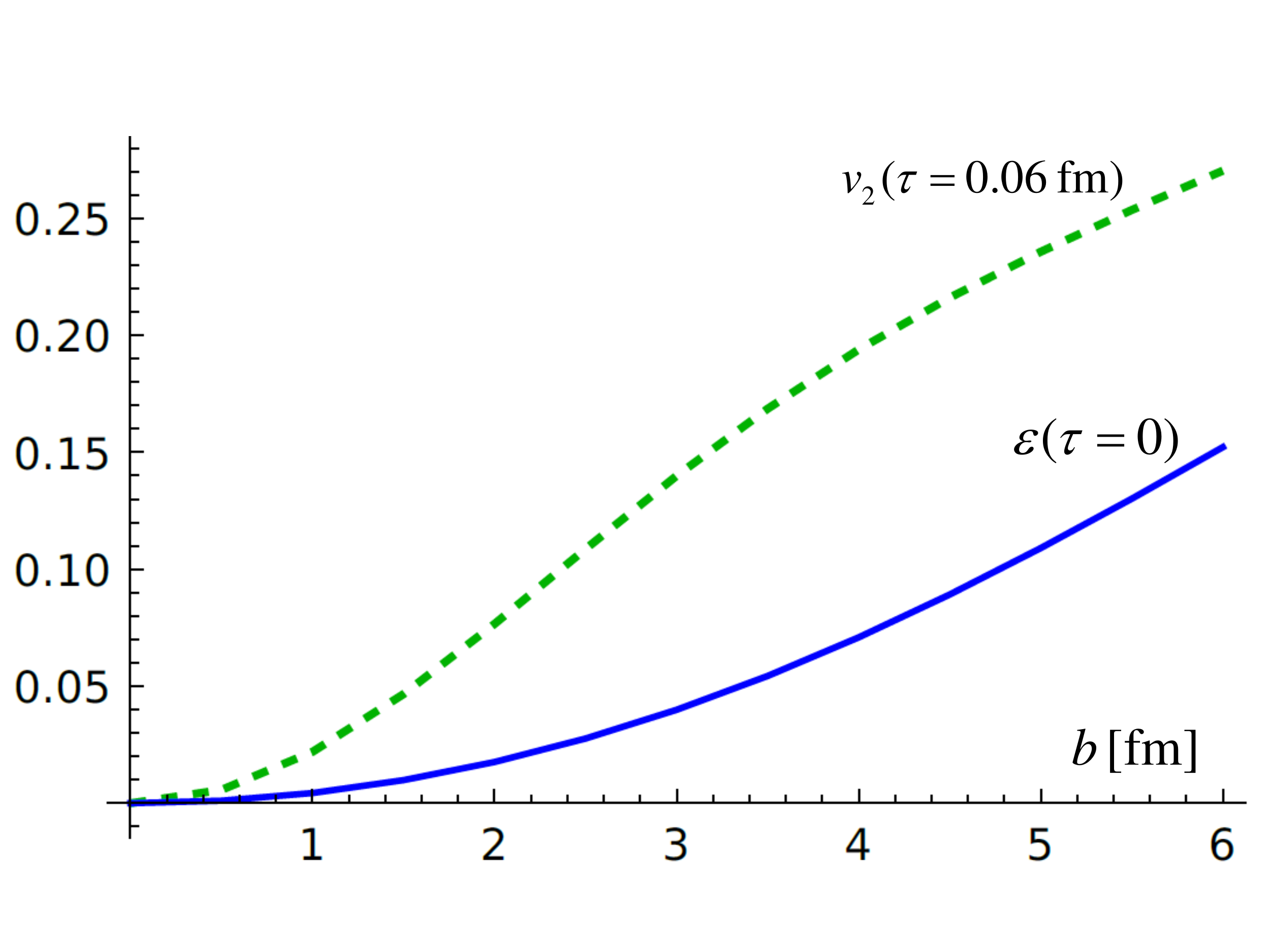}
\end{center}
\vspace{-15mm}
\caption{$v_2$ at $\tau=0.06$ fm and $\varepsilon$ at $\tau=0$ versus impact parameter.}
 \label{fig-v2-and-e2}
\end{figure}

\begin{figure}
\begin{center}
\includegraphics[width=10.5cm]{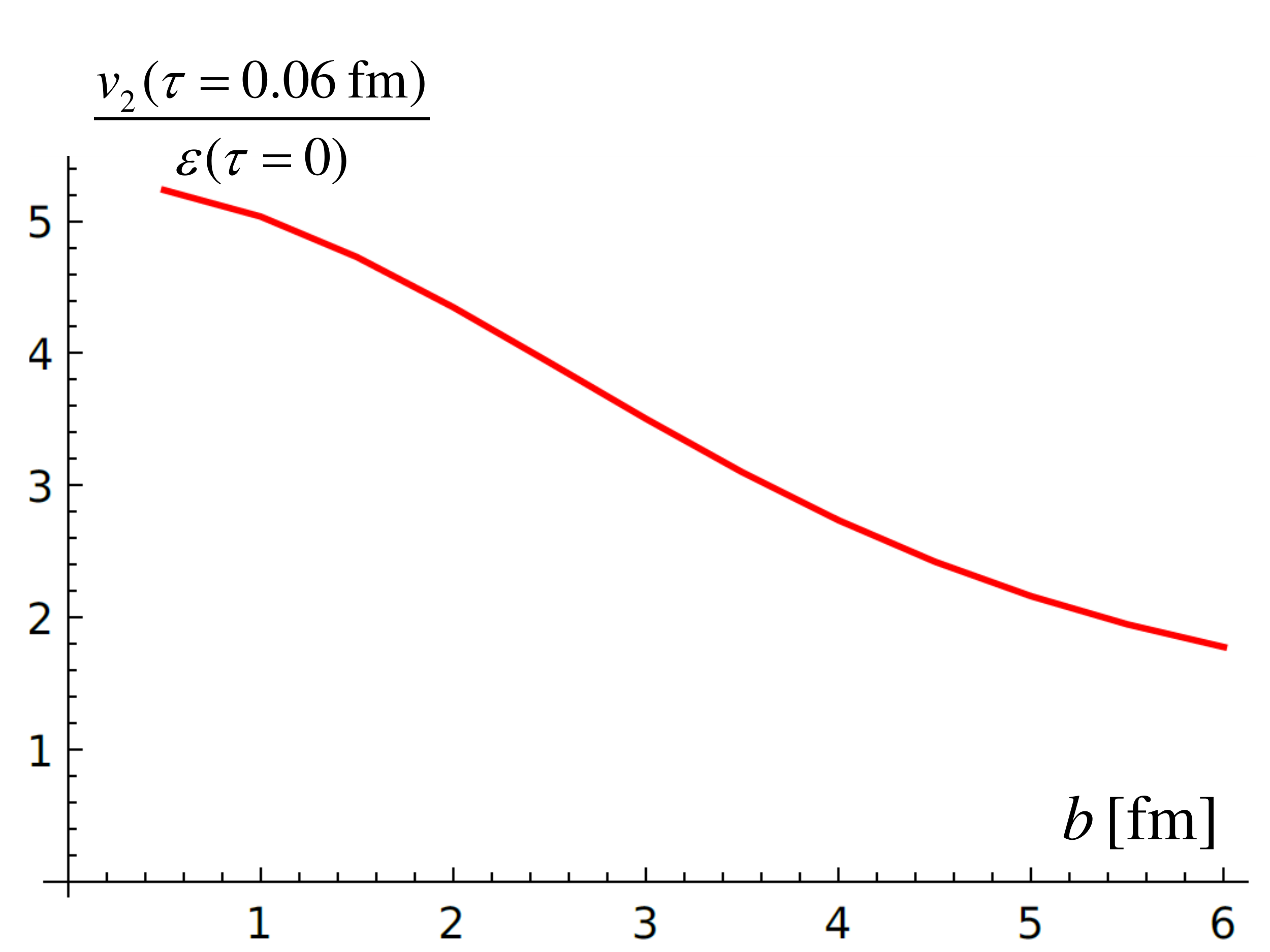}
\end{center}
\vspace{-7mm}
\caption{Ratio of $v_2$ at $\tau=0.06$ fm over $\varepsilon$ at $\tau=0$ versus impact parameter.}
 \label{fig-v2:e2}
\end{figure}

It is interesting to consider the dependence of the glasma elliptic flow on the system's initial eccentricity, which in turn depends on impact parameter. We show in Fig.~\ref{fig-v2-and-e2} the coefficient $v_2$ at $\tau=0.06$ fm computed at seventh order of the proper time expansion and the initial eccentricity, both as  functions of impact parameter. In Fig.~\ref{fig-v2:e2} the coefficient $v_2$ is divided by the initial eccentricity $\varepsilon$. One sees that the relative change in $v_2$ when the impact parameter grows from 1 fm to 6 fm is much greater than the relative change in the ratio $v_2/\varepsilon$. This behaviour indicates that  the initial spatial asymmetry of the glasma is transmitted to the momentum asymmetry of the system, which mimics the behaviour of hydrodynamics. 

We comment that in Ref. \cite{Carrington:2021qvi} we performed a similar analysis (at lower order in the proper time expansion), but the coefficient $v_2$ calculated for $\tau = 0.04$~fm was compared, not with the initial value of $\varepsilon$, but with its value at the same proper time $\tau = 0.04$~fm. Since the initial value of $\varepsilon$ differs by less than 1\% from its value at $\tau \lesssim 0.04$~fm, the numerical results and the conclusions we draw from them are unaffected by this issue.

\subsection{Angular momentum}
\label{sec-angular}

\begin{figure}
\begin{center}
\includegraphics[width=8.9cm]{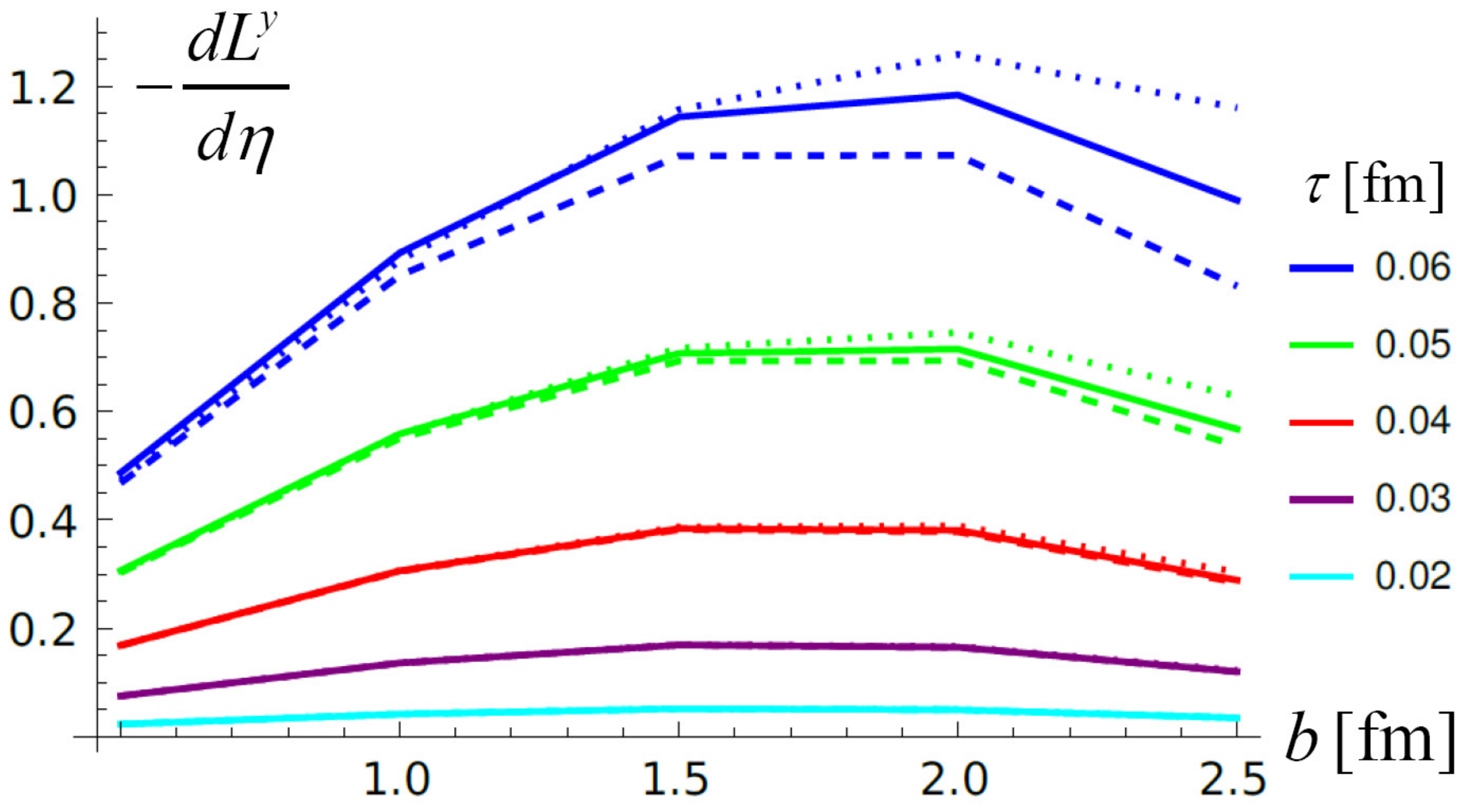}
\includegraphics[width=7.3cm]{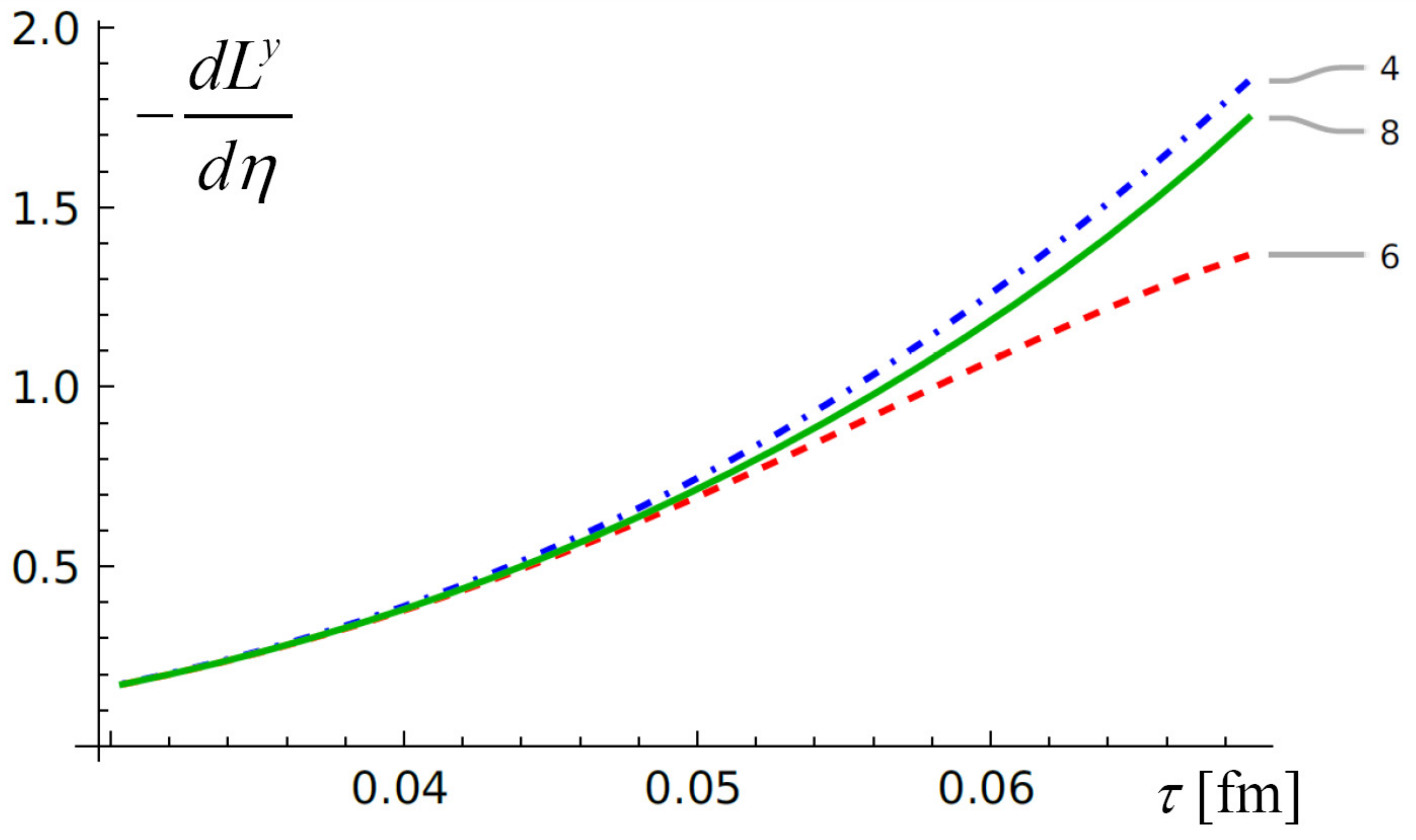}
\end{center}
\vspace{-6mm}
\caption{Left panel: angular momentum versus impact parameter at different times to fourth (dotted lines), sixth (dashed lines) and eighth (solid lines) order of the proper time expansion. Right panel: angular momentum versus time at different orders in the $\tau$ expansion with $b=2.0$ fm. }
\label{fig-L-time}
\end{figure}

A system of relativistic heavy ions colliding at a finite impact parameter has initially a huge angular momentum which is perpendicular to the reaction plane. The value of the initial angular momentum carried by the nucleons which will participate in the collision is of order $10^5$ at maximum RHIC energies \cite{Gao:2007bc,Becattini:2007sr} and even larger at LHC energies. We would like to know how much of this initial angular momentum is transferred to the glasma that is produced in the collision. Since the glasma in our approach is boost invariant, we cannot compute the total angular momentum of the system which, strictly speaking, extends in rapidity from minus to plus infinity. Instead we compute the angular momentum per unit rapidity which can be obtained from the formula \cite{Carrington:2021qvi}
\be
\label{L-def-4}
\frac{dL^y}{d\eta} = -\tau^2 \int d^2\vec R \, R^x T^{01} .
\ee
In the first graph in Fig.~\ref{fig-L-time} we show the glasma angular momentum at five different impact parameters from $b=0.5$ fm to $b=2.5$ fm and at five different proper times from $\tau = 0.02$ fm to $\tau = 0.06$ fm. We displace the ion moving in the positive $z$-direction a distance $b/2$ in the positive $x$-direction, and the ion that is moving in the negative $z$-direction is shifted the same amount in the negative $x$-direction. The collision therefore produces angular momentum in the negative $y$-direction. The results are obtained at fourth, sixth and eighth order of the proper time expansion. The second graph in Fig.~\ref{fig-L-time} shows the angular momentum computed at these three orders as a function of $\tau$. One sees that the eighth order results are very close to those of sixth order for $\tau \lesssim 0.06$ fm, which means that the time interval under consideration is well within the radius of convergence of the expansion. 

The integral over $R^x$ in Eq.~(\ref{L-def-4}) is taken up to $R^x_{\rm max} = 5.9$ fm. We note that the dominant contribution to the angular momentum comes from the parts of the nuclei that are farthest from the collision centre, with respect to which angular momentum is calculated. These are the regions where the gradient expansion we use is to be trusted the least. Our results for the angular momentum therefore do depend on the upper limit of the integral, and should only be considered order of magnitude estimates for the glasma angular momentum. In Ref. \cite{Carrington:2021qvi} we give a more detailed analysis of the extent to which our results for the angular momentum of the glasma depend on the integration region that is used to do the calculation.  It is important to note that the physical significance of our result is unaffected by these considerations. 
Our eighth order results confirm and reinforce our earlier finding that only a small fraction of the very large angular momentum of the incoming nuclei is transferred to the glasma.

\subsection{Jet quenching}
\label{sec-jet}

We study jet quenching using a Fokker-Planck equation of hard probes in a system populated with strong chromodynamic fields \cite{Mrowczynski:2017kso}
\be
\label{F-K-eq}
\Big({\cal D} - \nabla_p^\alpha  X^{\alpha\beta}(\vec{v}) \nabla_p^\beta - \nabla_p^\alpha  Y^\alpha (\vec{v}) \Big) n(t, \vec{x},\vec{p}) = 0,
\ee
where $n(t, \vec{x},\vec{p})$ is the distribution function of hard or heavy partons, $\vec{v}=\vec{p}/E_p$ is the parton's velocity, $\vec{p}$ is the momentum and $E_p$ is the energy. The parton's four-momentum is assumed to be on mass-shell and ${\cal D} \equiv  \frac{\partial}{\partial t} + \vec{v}\cdot \nabla $ is the substantial derivative. The tensor  $X^{\alpha\beta}(\vec{v})$ is 
\be
\label{X-def}
X^{\alpha\beta}(\vec{v}) \equiv
\frac{1}{2N_c} \int_0^t dt' \: {\rm Tr}\big[ \big\langle \mathcal{F}^\alpha(t, \vec{x}) 
\mathcal{F}^\beta \big(t-t', \vec{x}-\vec{v} t'\big) \big\rangle \big] ,
\ee
where ${\cal{\vec F}} (t,\vec{x}) \equiv g \big(\vec{E}(t,\vec{x}) + \vec{v} \times \vec{B}(t,\vec{x})\big)$ is the Lorentz color force and $g$ is the coupling constant. The chromoelectric $\vec{E}(t,\vec{x})$ and chromomagnetic $\vec{B}(t,\vec{x})$ fields are given in the fundamental representation of the SU($N_c$) group. The vector $Y^\alpha(\vec{v})$ can found from the relation
\be
\label{XY}
Y^\alpha(\vec{v}) = \frac{v^\beta}{T} X^{\alpha\beta} (\vec{v}) ,
\ee
where $T$ is the temperature of an equilibrated quark-gluon plasma that has the same energy density as the glasma. The collisional energy loss $dE/dx$ of a high-energy parton traversing the glasma and the momentum broadening coefficient $\hat{q}$ which determines a hard parton's radiative energy loss are determined by the tensor $X^{\alpha\beta}(\vec{v})$ as 
\ba
\label{e-loss-X-Y}
\frac{dE}{dx} &=& - \frac{v}{T} \frac{v^\alpha v^\beta}{v^2} X^{\alpha\beta}(\vec{v}), 
\\
\label{qhat-X-T}
\hat{q} &=&  \frac{2}{v} \Big(\delta^{\alpha\beta} - \frac{v^\alpha v^\beta}{v^2}\Big) X^{\alpha\beta}(\vec{v}).
\ea

We mention that the field correlators in equation (\ref{X-def}) are non-local and consequently they are not gauge invariant.
In principle this problem could be fixed by inserting a link operator between the two fields but practically this procedure is
difficult to realize. In Ref. \cite{Carrington:2021dvw} (see equations (26-28)) we gave an argument that the effect of omitting this link operator is probably not very large in our calculation. The key observation is that due to the short time interval the link operator is not much different from unity.  The generalization of our method to a gauge invariant formulation is an important open question that we intend to return to in a future publication. 

The correlators of gauge potentials that we have calculated using the small $\tau$ expansion provide the correlators of chromoelectric and chromomagnetic fields which determine the tensor (\ref{X-def}) and in turn give $dE/dx$ and $\hat{q}$. In our earlier works \cite{Carrington:2022bnv,Carrington:2021dvw} calculations were done to fifth order in the proper time expansion. We present below the results of our  latest seventh order calculations. We consider only the case where the ultra-relativistic hard probe moves perpendicularly to the beam axis, with $v=v_\perp=1$, and we note that this means the energy loss gets contributions only from even orders in the $\tau$ expansion. In Ref. \cite{Carrington:2022bnv}, working at fifth order in the proper time expansion, we studied the dependence of the momentum broadening parameter on the magnitude and direction of the probe's velocity. 
The size of the momentum broadening parallel and perpendicular to the beam was studied in \cite{Ipp:2020mjc}. 
\begin{figure}
\begin{center}
\includegraphics[width=11cm]{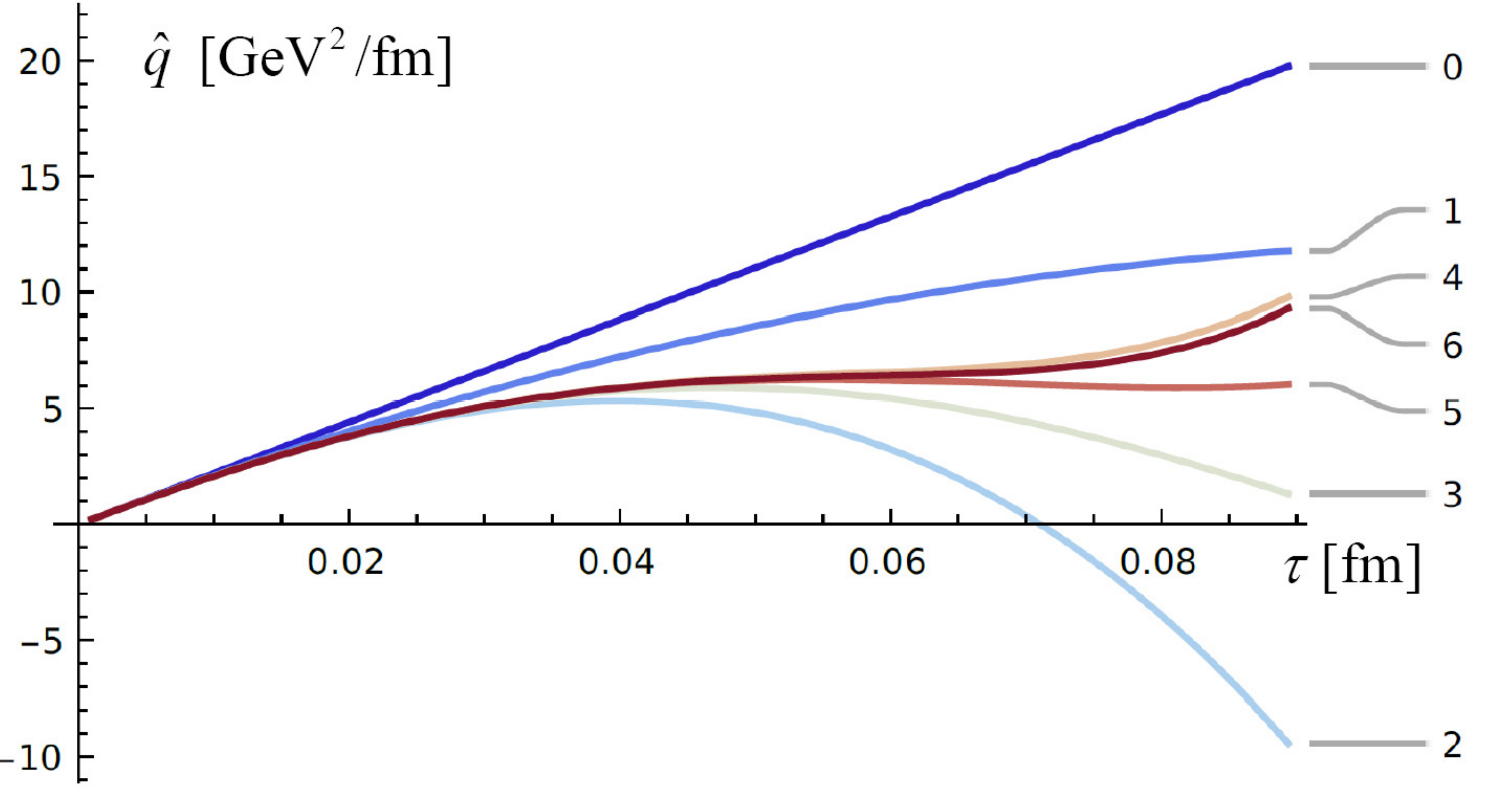}
\end{center}
\vspace{-9mm}
\caption{The momentum broadening coefficient $\hat q$ as a function of $\tau$ at different orders of the proper time expansion. The seventh order result cannot be seen because it lies directly under the sixth order one.}
\label{fig-qhat-time} 
\end{figure}

In Fig.~\ref{fig-qhat-time} we show the momentum broadening coefficient $\hat q$ as a function of $\tau$ at different orders of the proper time expansion up to sixth order. The seventh order result is not seen because it lies directly under the sixth order one, which indicates that the proper time expansion converges to high accuracy up to approximately $\tau = 0.08$ fm. We comment that for $\hat q$ different orders in the expansion do not exhibit a simple behaviour of alternately increasing and decreasing the result, as seen in quantities obtained from the energy-momentum tensor, for example in figures \ref{figure4-repro} and \ref{fig-P-time}. 

Recently calculations of $\hat q$ have been done using a kinetic theory description of an anisotropic quark-gluon plasma \cite{Boguslavski:2023alu}, which is valid between the very early times where the glasma exits and the onset of hydrodynamics. 
The results of this calculation smoothly connect the two regimes, and support the idea that the pre-equilibrium phase plays an important role in jet quenching.

\begin{figure}
\begin{minipage}{81mm}
\centering
\includegraphics[width=8.7cm]{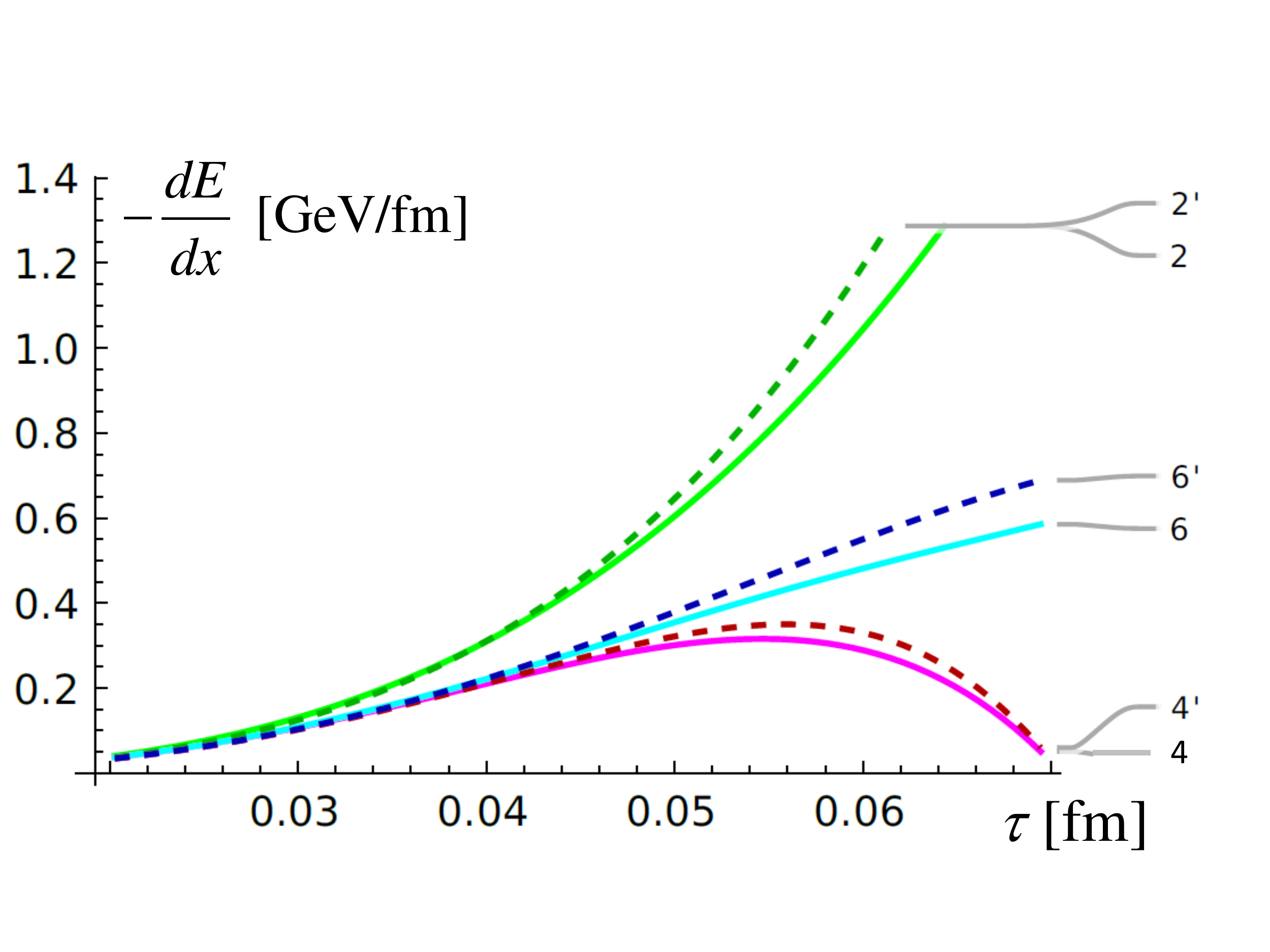}
\end{minipage}
\begin{minipage}{81mm}
\centering
\vspace{5mm}
\includegraphics[width=7.3cm]{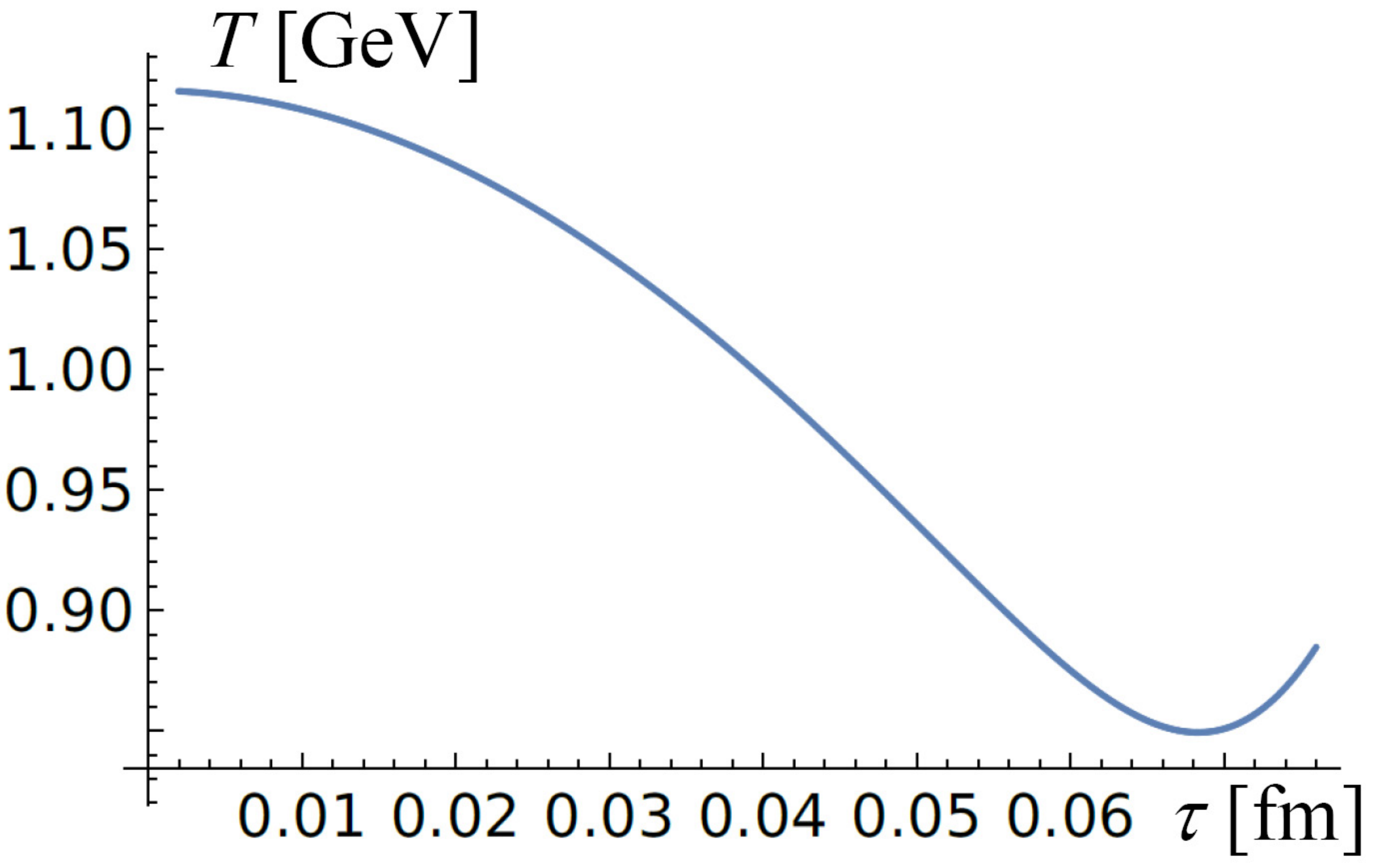}
\end{minipage}
\vspace{-7mm}
\caption{Left panel: the energy loss as a function of $\tau$ at different orders of the proper time expansion up to sixth order. The dashed lines are divided by an effective temperature obtained from the glasma energy density (see text for details) and the lighter coloured solid curves are made with $T=1$ GeV. Right panel: the effective temperature as a function of the proper time, determined by comparison with an equilibrium system with the same energy density. }
\label{fig-eloss-time}
\end{figure}

The collisional energy loss $-dE/dx$ of a hard parton moving with $v=v_\perp=1$ is shown as a function of $\tau$ in the first graph in Fig.~\ref{fig-eloss-time}. In order to calculate $dE/dx$ we need the temperature $T$ of an equilibrated quark-gluon plasma whose energy density is the same as the energy density of the glasma. Using the formula for the energy density of an equilibrium noninteracting quark-gluon plasma, the effective temperature of the glasma can be estimated from the glasma energy density. The temperature obtained from the eight order energy density is shown in the second graph in Fig.~\ref{fig-eloss-time}, and the curves for which the order of the expansion is indicated with a prime are obtained using this temperature. The lighter curves on the left side plot are obtained using a constant value $T=1$ GeV, which is not as well motivated from a physics point of view but has the advantage of not mixing the dependence of the two calculations, $-dE/dx$ and ${\cal E}$, on the proper time expansion. The figure shows that the results at different orders converge well up to $\tau \approx 0.06$ fm for both calculations. We comment that the results in our previous paper were obtained using an effective temperature calculated from the sixth order energy density (instead of our current eighth order result) and for this reason the region of validity of the expansion was significantly smaller.

In our earlier works \cite{Carrington:2020sww,Carrington:2022bnv,Carrington:2021dvw} the momentum broadening coefficient $\hat q$ and the collisional energy loss $dE/dx$ were computed only in the simple case of a homogeneous glasma, which means that the incoming nuclei were assumed to be infinitely extended and homogeneous in the plane transverse to the beam direction. A realistic modeling of jet quenching in relativistic heavy-ion collisions requires treating nuclei as finite objects of varying density. We have generalized our previous calculations of $\hat q$ and $dE/dx$ so that the glasma under consideration is produced in collisions of finite nuclei with a Woods-Saxon density distribution. The field correlators are computed using a first order gradient expansion. The correlator $\langle \beta_n^i(\vec x_\perp) \, \beta_n^j(\vec y_\perp) \rangle$ is expanded around $\vec R \pm \vec b/2$ where $\vec R = \frac{1}{2}(\vec x_\perp + \vec y_\perp)$ and only the first two terms of the expansion are included. We note that the correlator $\langle \beta_n^i(\vec x_\perp) \, \beta_n^j(\vec y_\perp) \rangle$  is independent of $\vec R$ if the system is  translationally invariant in the transverse plane.
 
\begin{figure}
\begin{center}
\includegraphics[width=8.3cm]{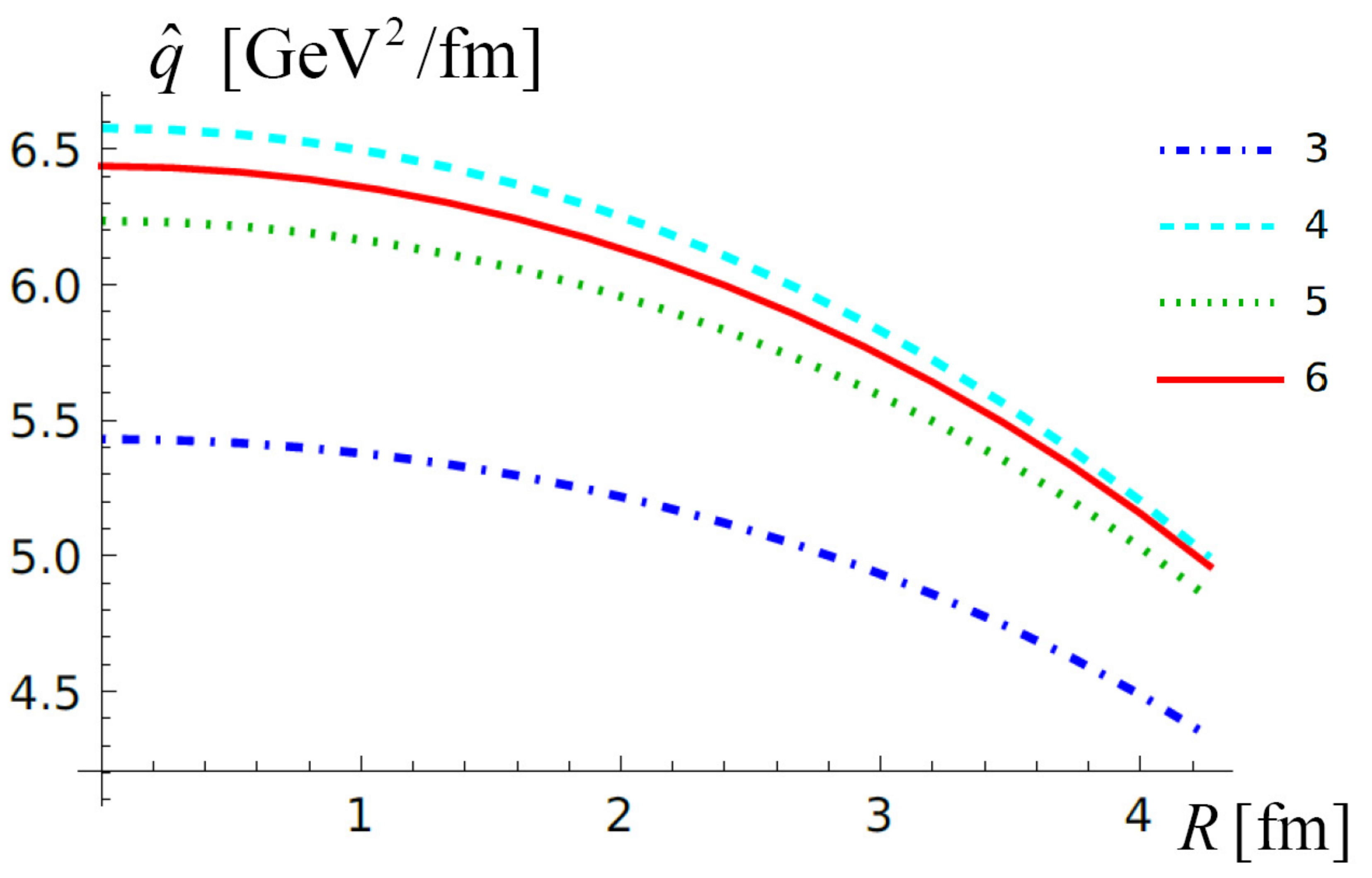}
\includegraphics[width=7.9cm]{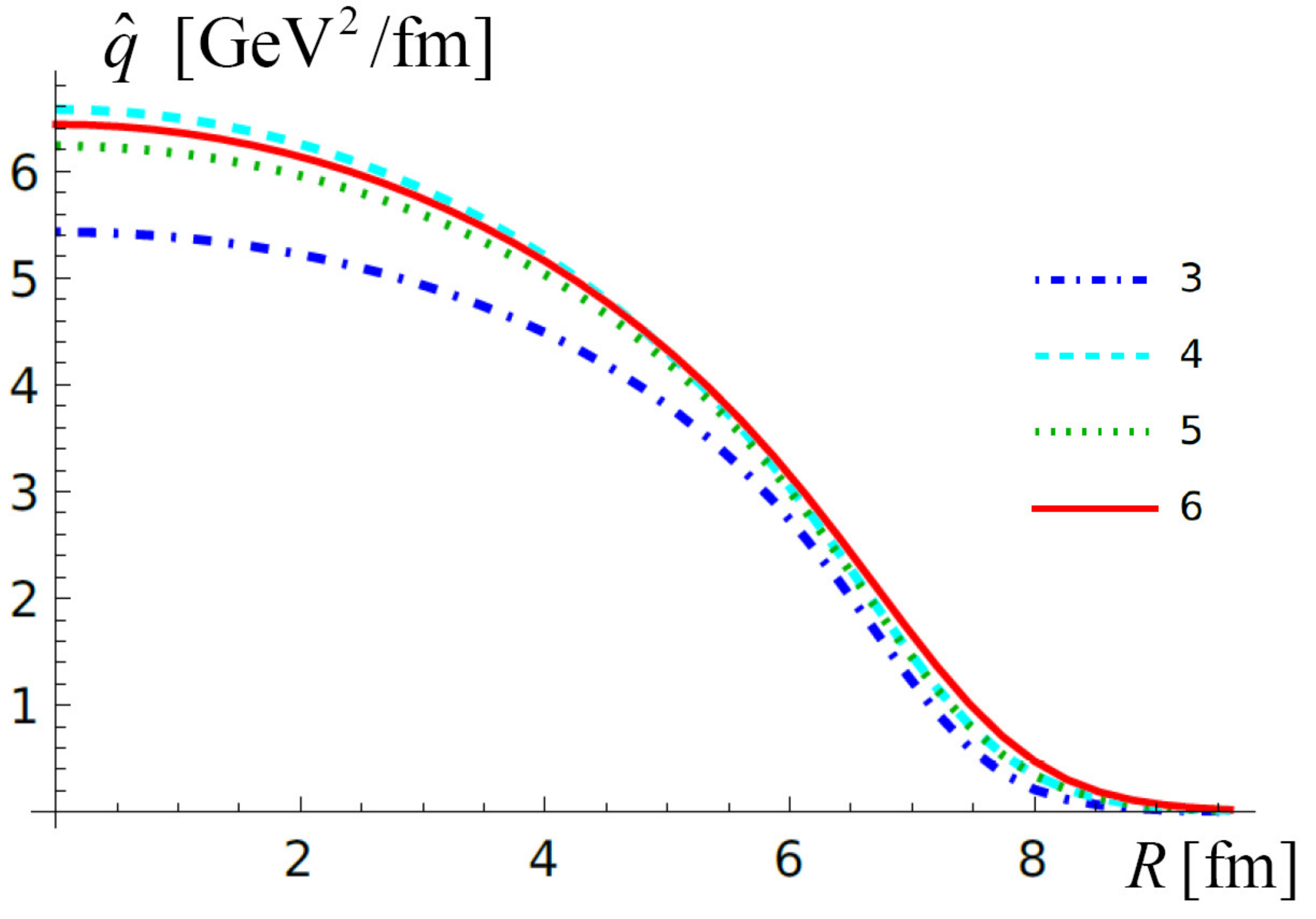}
\end{center}
\vspace{-7mm}
\caption{$\hat q$ versus $R$ for central collisions at $\tau=0.04$ fm and different orders of the proper time expansion from 3 to 6. The left panel shows a close up of the region $R<4$ fm.}
\label{fig-qhat-R}
\end{figure}

Figure \ref{fig-qhat-R} shows $\hat q$ versus $R$ for central collisions ($b=0$) at $\tau=0.06$ fm at different orders of the proper time expansion up to sixth order. 
The calculation is done for a lead nucleus with radius $R_A= 7.4$ fm. In the outer part of the system the charge density drops rapidly to zero and the gradient expansion is not reliable. In the region $0<R\lesssim 4$ fm, which covers most of the glasma's volume, $\hat q$ depends on $R$ only weakly.  This means that the assumption of translation invariance in the transverse plane is valid to good accuracy in this domain.

\section{Summary and conclusions}
\label{sec-conclusions}

We have extended and generalized our earlier calculations \cite{Carrington:2020ssh,Carrington:2021qvi,Carrington:2020sww,Carrington:2022bnv,Carrington:2021dvw}. The characteristics of glasma derived from the energy-momentum tensor, which were studied up to fifth or sixth order in the proper time expansion, have been calculated to eighth order for all quantities that can be obtained from the energy momentum tensor. The calculations of the transport coefficients associated with collisional energy loss and momentum broadening have been extended from fifth to seventh order. The calculations presented in this paper enlarge the interval of $\tau$ where our results are reliable to $0 < \tau \lesssim 0.07~{\rm fm}$. Our results have allowed us to study in detail the isotropization of the system and its radial flow. We have improved our analysis of the Fourier coefficients of azimuthal distribution, which reveals that the evolution of the glasma has some features that mimic hydrodynamics. We have also studied the transfer of the angular momentum from valence quarks to glasma. 

Our calculations of the jet quenching parameters $\hat q$ and $dE/dx$, which were obtained previously up to fifth order, have been now extended to seventh order and they are significantly more precise. These calculations have also been generalized and made more physically relevant in an important way.  In our original approach, transport coefficients were calculated with the simplifying assumption that the incoming nuclei were translationally invariant in the plane transverse to the beam direction, and consequently the glasma was homogeneous. In this work we have included some of the effects of nuclear structure by representing the charge density functions of the nuclei with a Woods-Saxon distribution, and the effect of varying nuclear density has been taken into account using a gradient expansion up to first order. We note that the momentum broadening and collisional energy loss calculations are considerably more difficult than those of the energy momentum tensor because the latter requires one-point correlators while two-point correlators are needed to calculate transport coefficients. Our results show that the density dependent parameters $\hat q$ and $dE/dx$ depend only weakly on position in the transverse plane, except in the outer region of the glasma system. These results justify the assumption of approximate translation invariance in the transverse plane which is required for the applicability of the Fokker-Planck approach that we have developed. 

The results presented in this paper confirm and reinforce our earlier findings from \cite{Carrington:2020ssh,Carrington:2021qvi,Carrington:2020sww,Carrington:2022bnv,Carrington:2021dvw}, extend the radius of convergence of the proper time expansion, and verify that the method we have developed to calculate transport coefficients is compatible with realistically varying nuclear charge densities. This work establishes more firmly the validity of proper time expansions as a calculational method to study glasma properties, and motivates future projects using the methods we have developed. Specifically, the density dependent transport coefficients we have obtained allow for a more realistic modeling of jet quenching in nuclear collisions. 

\section*{Acknowledgments}

This work was partially supported by the Natural Sciences and Engineering Research Council of Canada under grant SAPIN-2017-00028 and the National Science Centre, Poland under grant 2018/29/B/ST2/00646. 


\end{document}